\documentclass[a4paper,11pt]{article}
\usepackage{jheppub} 
\usepackage{lineno}
\usepackage{float}
\usepackage{multirow}

\title{\boldmath Description of $Z$ Boson Mass and $p_T$ Spectrum at LHC Using Leading-Order Event Generators at $\sqrt{s} = 13.6$ TeV}

 \author{Dharmender}
 \author{and Bibhuti Parida}

\affiliation{Department of Applied Physics,\\
Amity Institute of Applied Sciences,\\
Amity University Uttar Pradesh, Noida, India}

\emailAdd{bparida@amity.edu}

\abstract{In this paper, we report a study of $Z$ boson kinematic distributions in the $Z$+ 1-jet events from proton-proton ($pp$) collisions at $\sqrt{s}$ = 13.6 TeV at the Large Hadron Collider (LHC) using Leading-Order (LO) event generators. Three LO event generators, namely Pythia8, Herwig7 and Sherpa2, are explored with the implementation of appropriate Parton Shower (PS) and Matrix Element (ME) corrections. The $Z$ boson is reconstructed from oppositely charged di-lepton pairs ($\mu^+\mu^-$ and $e^+e^-$), and the associate leading jet is reconstructed with the anti-$k_T$ algorithm with radius $r$=0.4. The di-lepton invariant mass, $M_{\ell^+\ell^-}$, and the transverse momentum of the reconstructed $Z$ boson, $p_{T}^Z$, distributions are extensively studied by comparing three of the Monte Carlo (MC) predictions. In addition to this, we also explore several other kinematic distributions associated with the candidate lepton and the jet to distinguish the modelling of the LO event generators in depth. Furthermore, various showering modules within these event generators are explored for $Z$ + 1-jet production. While the three of these MC frameworks reproduce the $Z$ + 1-jet kinematics similarly, certain discrepancies in their predictions are identified in some variables. We analyzed them using different statistical tests. The best match of the $p_{T}^Z$ spectra was found between the Pythia8 and Sherpa2 results.\\

}

\begin{document}
\maketitle
\flushbottom
\pagebreak
\section{Introduction}
\label{sec:intro}
The $Z$ boson production in association with jets is a common background process in the high-energy hadronic collisions. This process is highly sensitive to several Standard Model (SM) processes, including the study of the Higgs boson and top-quark properties, as well as searches for new physics such as dark matter and supersymmetry, at the LHC. The $Z$+jet process is one of the prominent example of hard-scattering process which yields a very distinct experimental signature at the hadron colliders. In terms of Monte Carlo (MC) event generators, this process serves as an important benchmark for studying the perturbative QCD descriptions that accompanies the partons in a hard-scattering process. This, in turn, reveals the underlying modeling and implementations in the simulation frameworks. Thus a thorough understanding of this process is crucial for testing the SM predictions as well as searches for physics Beyond the SM (BSM).

Numerous theories and experiments have been put forward to precisely predict the outcomes of this process at different colliding energies and with the help of advanced perturbative QCD calculations, we can explore the predictions upto the level of Next-to-Next Leading order (NNLO). With the help of event generators such as Pythia8\cite{Bierlich:2022pfr}, Herwig7\cite{Bahr:2008pv} and Sherpa2\cite{Sherpa:2019gpd}, which perform the calculations at the Leading Order (LO) approximations, we can simulate the complex picture of the high energy hadronic collisions. With appropriate theoretical models and assumptions, the hard scattering process associated with $Z$ boson production can be isolated by applying the corresponding mathematical function derived from the Feynman diagrams. The MC simulations such as Pythia8, Herwig7 and Sherpa2 use different assumptions and techniques to describe the kinematics of the whole process upto a certain accuracy level. Generally, the approximation of higher-order real-emission corrections to the hard scattering is usually done by Parton Shower (PS) methods\cite{Frixione:2007vw} in MC simulations which is efficient in describing the QCD radiations in soft $p_T$ regions. However, being a collinear approximation it is unreliable at precisely describing the hard region of the phase-space especially in regions where the phase-space is widely stretched. 

 Focusing on a precise analysis, there are many options to provide theoretical accuracy to the simulations. In our analysis, we use the default Matrix Element (ME) corrections\cite{Catani:2001cc, 2009_paper} to correct emissions originating from the PS in various event generators. These corrections use information derived from exact ME calculations such that at high $p_T$, the interaction is characterized by the precise first-order perturbative calculation of QCD whereas in the low $p_T$, region, where multiple radiation and non-perturbative effects of QCD are dominant, the production kinematics are described by the PS.


In this paper, we explored different kinematic distributions of the $Z$+ 1-jet process with the leptonic decay mode of the $Z$ boson, as modeled by the most popular event generators Pythia8, Herwig7 and Sherpa2, which treat the processes at LO approximations. We investigated the higher-order corrections on the kinematic distribution of the $Z$ boson, associated jet and the leptons using the ME corrections as implemented in Pythia8, Herwig7 and Sherpa2 and compare their predictions.

\section{Description of the \textit{Z} Boson  Kinematics}
The $Z$ boson kinematic distributions are well-described by the theoretical predictions based on the SM where the dilepton invariant mass distributions ($M_{\ell^{+}\ell^{-}}$) peak at 91.18 GeV/$c^2$ and the transverse momentum ($p_{T}^{Z}$) peaks at low $p_T$ and falls off rapidly at higher $p_T$ region. The peaks of the distributions are sensitive to the production mechanism of the $Z$ boson, while the tails of the distributions are sensitive to higher-order QCD effects\cite{Vesterinen:2008hx,Collins:1984kg,Odaka:2009qf}.

In terms of MC simulations, the shape of the $p_{T}^{Z}$ spectrum describes the overall processes involved in the production and decay, and tests the modelling of showering and accuracy of QCD implemented in these simulations, since non-zero $p_T$ is generated through radiation from the initial state partons. With few exceptions, the generic signatures of production and decay of new particles include a mixture of leptons and jets, which can dominant residuals of SM process such as $Z$+jet production at the LHC \cite{Campbell:2011zz}. The associated production with a $Z$ boson decaying to a lepton pair with jets can lead to final states with jets and missing energy. Thus, a thorough study of the kinematic distributions of this process is an essential aspect of experimental particle physics that describes the underlying physics of the particle collisions. The precise measurement of the $p_T^Z$ distribution, provides valuable insights of the dynamics of the production and decay of the $Z$ boson, which in turn sheds light on the fundamental interactions of the SM. Due to its clear experimental signature in leptonic decay modes, the $Z$ boson production has been extensively studied in $pp$ collisions at the CMS and ATLAS experiments at LHC\cite{CMS:2011wyd,CMS:2019raw,CMS:2022ubq,ATLAS:2014alx,ATLAS:2015iiu,ATLAS:2019zci}.

Since after the experimental validation of this massive vector boson from $p\overline{p}$ collisions at the UA1 and UA2\cite{UA2:1983mlz, UA2:1983tsx,Hansen:1984mw} experiments, CERN in 1983, the ATLAS and the CMS Collaborations at LHC are still looking for more precise measurements of the $Z$ boson in the new energy regime of $pp$ collisions. Through a precise study of the $Z$ boson kinematic description, investigating its invisible decay channels, couplings, as well as its interactions with other heavy particles, the ATLAS and CMS Collaborations are exploring any deviations from the SM framework that could potentially provide insights into new physics.

\section{MC Simulations}
\subsection{Event Generation}
MC events in this analysis were simulated using the Pythia8, Herwig7 and Sherpa2 multi-purpose event generators, which simulate the hadronic collisions at LO approximations. These are the most commonly used event generators in high energy physics, capable of describing the high energy particle collisions over the full range of energy scales possible in current experiments. With the help of these simulations, theoretical calculations for a specific process for example, $Z$ boson production from  $q$ and anti-quark $\overline{q}$ pair could be well described.
A brief description of these event generators are given below.

\begin{itemize}
    \item \textbf{Pythia8:} A general purpose event generator capable of describing lepton-lepton, lepton-hadron, and hadron-hadron collisions with configurable beam properties, such as beam energies and crossing angles, to simulate one or many SM processes. It offers different options for the user to control the parton level and hadron level processes including Initial State Radiation (ISR), Final State Radiations (FSR), MultiPartonic Interactions (MPI), Underlying Events (UE) and several other event generation options.
    
    \item \textbf{Herwig7:} The Hadron Emission Reactions With Interfering Gluons or Herwig is a general purpose event generator similar to Pythia8 in describing the high-energy lepton-lepton, lepton-hadron and hadron-hadron collisions with special emphasis on the accurate simulation of QCD radiation.
    
    \item \textbf{Sherpa2: } Sherpa2 is a multi-purpose MC event generator for the Simulation of High-Energy Reactions of PArticles capable of describing lepton-lepton, lepton-photon, photon-photon, lepton-hadron and hadron-hadron collisions, with an emphasis on a multi-leg matrix element approach. It offers a large variety of higher-order corrections for the precise study of processes with hard radiations and multijet processes.
\end{itemize}

The main difference in predictions in these event generators lies in the implementation of their PS algorithms, tuning parameters and hadronization modelling which affects the overall description of particle kinematics. In this study, we compare the predictions in modelling the kinematics of the reconstructed $Z$ boson, its decayed leptons and jets associated with its production.

\subsection{Shower Generation and Hadronization}
Even though the PS algorithms in Pythia8, Herwig7 and Sherpa2 do the same job of branching colored parent partons into daughter partons, they have different approach of implementing the parton showering. 
In Pythia8, the showering is done with the help of a QCD dipole shower\cite{Bierlich:2022pfr, Sjostrand:2004ef} with some modifications. The QCD picture consists of two quarks ($q$ and $\overline{q}$) having opposite and compensating colours forming a colour dipole and out of the two parents (dipole ends) $\Tilde{a}$ and $\Tilde{b}$, one is said to be the ``emitter", which has more energy to emit daughter partons while the other, the ``recoiler” or spectator ensures the four-momentum conservation.

If we consider the branching the mother parton $a$ $\rightarrow$ $b$ +$c$  daughter partons, the differential probability of parton $a$ to branch is given by:
\begin{equation}
 \mathrm{d} \mathcal{P}_a\left(z, Q^2\right)=\frac{\mathrm{dQ}^2}{Q^2} \frac{\alpha_{\mathrm{s}}\left(Q^2\right)}{2 \pi} \sum_{b, c} P_{a \rightarrow b c}(z) \mathrm{d} z.
\end{equation}

 where $z$ is the momentum fraction of parton $b$ carried from parent $a$ and $(1-z)$ is the momentum fraction occupied by $c$ parton and $\alpha_\mathrm{s}$ is the coupling constant\cite{Bierlich:2022pfr}. The PS algorithms are formulated as an evolution of virtualities of the branching partons and the term $Q$, is an important aspect in every parton shower modelling that differentiates parton shower implementations in different event generators. Depending on this evolution scale, the branching is ordered in the form of transverse momentum, mass, angle between the branched products etc. 


The default showering in Herwig7 uses a angular-ordered shower (`$q$-tilde shower')\cite{Gieseke:2003rz, Bellm:2019zci} approach in which every coloured leg of the hard process are considered to be a shower progenitor, which branches into daughter partons and the kinematics of the daughter particles are determined by the set of vectors those are assigned to each of these coloured legs via the Sudakov basis \cite{Gieseke:2004tc} as: 
\begin{equation}
q_i = \alpha_i p + \beta_i n + q_{\perp i}. 
\end{equation}

where the vector $p$ is equal to the momentum of the coloured leg considered as the `shower progenitor' generated by the prior simulation of the hard scattering process. The reference vector $n$ is a light-like vector that satisfies $n\cdot p$ $>$ $m^2$ (and is usually chosen in such a way that it is anticollinear to $p$ in the frame where the shower is generated, maximizing $n$ · $p$). The $q_{\perp i}$ vector gives the remaining components of the momentum, transverse to $p$ and $n$ \cite{Bahr:2008pv}. 

In the Sherpa2 event generator, the default parton showering is based on the Catani-Seymour (CS) dipole factorisation \cite{CS_1, Catani:2002hc}. This CS shower or CSS \cite{Schumann:2007mg} provides a systematic way of calculating the probabilities of emitting partons at different energies and angles where the shower starts from the hard scattering process and then evolves the partons through successive emissions. The emissions are simulated based on the probabilities calculated using the CS dipole formalism. The branching of the parton in the shower is ordered in the form of a variable in the shower evolution. The choice of this evolution scale is different in case of Pythia8, Herwig7 and Sherpa2, which can give different outcomes for the same process. Herwig7 implements the coherent showering with the angular ordering \cite{Gieseke:2003rz} of PS emissions for a better treatment of the soft gluon emissions whereas in case of Pythia8 and Sherpa2, the PS emissions are $p_T$ ordered \cite{Sjostrand:2004ef, Sherpa:2019gpd} with which the issue of vetoing splittings with scales that are larger than the scale set by the hard process is handled accordingly.


The above description of the branching leaves coloured particles in the final-states, which is not practical. To tackle this issue, different event generators implement different approaches of confining these colored final-state particles and hadrozination of colorless hadrons from these colored particles. Two most commonly used phenomenological approaches are: String model \cite{Andersson:1983ia, Sjostrand:1984ic} and Cluster Model \cite{hadron_cluster}. In the former approach, a linear rising potential between a quark $q$ and anti-quark $\overline{q}$ pair is considered, which increases with distance, caused by a coloured string or flux tube due to self interaction of gluons. Formation of quark $q$ and anti-quark $\overline{q}$ pairs take place when this string is stretched or has enough energy, which eventually stops when only an on-mass-shell hadron is left in the end. In the latter approach, nearest quarks, coming from splitting of gluons into light quark-antiquark or diquark-antidiquark pairs, are combined together to form color singlet clusters and the decayed hadrons are decided from the mass of these clusters. Pythia8 uses the String Model whereas Herwig7 uses the Cluster model \cite{Bahr:2008pv} for the hadronization. In Sherpa2, the default hadronization model (AHADIC++) is based on the cluster fragmentation model with some modifications \cite{hadron_cluster}.

\subsection{Matrix Element Corrections}
The PS in these event generators uses a soft radiation model to simulate the emission of gluons and quarks from the incoming and outgoing partons in high-energy particle collisions. This model is based on the assumption that the soft radiation is emitted independently of the underlying hard scattering process, and can therefore be factorized from the hard process. However, this assumption breaks down for hard or complex processes that are not well-described by the soft radiation modelled in the shower. In these cases, the PS simulation alone may not be able to accurately model the kinematics of the final state particles. For an improved description, the PS is strategically combined with the ME calculations, which includes all possible feynmann diagrams for the process, including virtual and real corrections. The ME calculations produce a set of differential cross sections for the final state particles, which describe the probability of producing each final state configuration.

 In the event generators used in this study, ME strategically corrects the hard emissions coming from the PS approach. This technique generates events by first calculating the ME for the hard process, then simulating the radiation of additional partons using the PS. For this, the phase-space is divided into two well defined regions and the emissions are generated using the above approaches. The corrections are implemented in such a way that the double counting and dead zones are avoided by preventing any overlap between the two phase-space regions and together covering the entire phase-space. Merging scale is the one that handles the border or the separation of the soft-hard $p_T$ regions. 

 For this analysis, we used the default ME correction which is implemented in Pythia8, that corrects the hardest emission from the initial (spacelike shower) and final state (timelike shower) emissions\cite{Bierlich:2022pfr}. Since the PS is $p_T$ ordered in Pythia8, it is usually the first emission which needs to be corrected. Thus, the first branching produced off the hard sub-process is corrected to the full ME at leading-order accuracy in this case. In case of Herwig7, there is a two step approach used for the hard and/or soft ME corrections implemented by default to carry out the re-summation in the parton showering. In the phase space region which is left uncovered by the Herwig7 PS (`dead-zones'), the exact ME for one additional parton is used for the correction with the successive shower (known as Hard ME correction)\cite{Bahr:2008pv, 2009_paper} and where the phase space region is occupied by the PS, the ME corrections are applied similar to the ME corrections in Pythia8, but the corrections has to be applied to the hardest emission from all of the PS emissions\cite{Seymour:1994df}, not the first emission (which is the case in Pythia8) and is referred as the `soft ME corrections'. In Sherpa2, the ME corrections are based on CKKW merging technique \cite{Catani:2001cc, Krauss:2002up} at LO which is labelled as MEPS@LO\cite{Hoeche:2009rj} for LO calculations which relies on truncated parton shower, i.e. the shower explicitly generating the Sudakov form factor for lines between reconstructed matrix–element-type emissions~\cite{Sherpa:2019gpd}.

\section{Analysis Procedure}
\subsection{Analysis Setup}
Latest versions of  Pythia8 (v8.310), Herwig7 (v7.2.3) and Sherpa2 (v2.2.15) with the default settings were used for the $pp$ collisions in our study. 
In Pythia8 event generator, the default Monash 2013 tune\cite{Skands:2014pea} for $pp$ collision is used whereas in Herwig7, H7.1-Default tune \cite{Bellm:2019zci} has been implemented which is the updated version of H7-UE-MMHT tune\cite{Bahr:2008pv} used in previous 7.0 versions. In Sherpa2, the parameters in the default tune are tuned with the data from LEP 1 and that is discussed in Ref.\cite{Sherpa:2019gpd}.  
For the Parton Distribution Function (PDF) choice, we use the internal PDF sets available in these event generators. The default NNPDF2.3 QCD with QED corrections (LO)\cite{Ball:2013hta, Bierlich:2022pfr} PDF set is used in Pythia8 whereas in Herwig7, the default CT14 PDF\cite{Dulat:2015mca, Bellm:2019zci} set is used. In Sherpa2, we considered the standard NNPDF 3.0 NNLO\cite{NNPDF:2014otw} PDF set.
Along with these MC event generators, fastjet package (v3.4.0)\cite{Cacciari:2011ma} was used for jet-clustering and jet finding in the high-energetic $pp$ collisions.
\subsection{Event Selection and Object Reconstruction}
\label{event_select}
MC events were generated for the $Z$ boson production in association with  1-jet from $pp$ collisions at center-of-mass energy of 13.6 TeV. A total of  1 milion events were simulated for the leptonic decay of $Z$ boson, particularly in the $e^{+}e^{-}$ and $\mu^{+}\mu^{-}$ decay channels, with minimum $\gamma$ contribution. The final state $Z$ boson decay candidates were selected by applying the following selection criteria.
\begin{itemize}
    \item In the leptonic decay channel of the $Z$ boson, we discard any lepton candidate which has transverse momentum ($p_T$) less than 25 GeV/c.
    \item Even though we have not used the detector simulation, we are considered the typical setup of LHC for finding the leptons within the central region i.e. $-2.5 < \eta < 2.5$.
    \item Anti-$k_{T}$ jets \cite{cacciari2008anti} with radius parameter $r$ = 0.4 is used in this analysis. For jet selection, we have considered the only jets which have $p_T >$ 30 GeV/c within the pseudo-rapidity range of -3.0 $< \eta < 3.0$.  

    \item To prevent any overlap between jets and leptons, a separation cut of $|\Delta R| > 0.4$ between the leading jet and the leptons is applied, where $\Delta R = \sqrt{(\Delta\phi)^{2}+ (\Delta\eta)^{2}}$ is the radius of cone around the lepton candidate.
    \item In order to prevent the contribution of the sub-leading jet on the leading jet, we imposed a constraint on the ratio ${p_{T}^{jet 2}}/{p_{T}^{Z}}$, ensuring it remains below 15\%.
    \item To ensure the leading jet and the $Z$ boson are back to back, we apply a delta phi cut of 2.94 radian (i.e, $|\Delta \phi |>$ 2.94 radian) between the $Z$ boson and the leading jet.
\end{itemize}
A summary of selections used in this analysis is described in Table 1.
\begin{table}[H]
    \centering
    \begin{tabular}{|c|c|c|}
    \hline
        \multicolumn{2}{|c|}{\textbf{ Requirement}} & \textbf{ Selection criteria} \\
        \hline
        \multirow{2}{*}{electrons} & Transverse momentum & $p_T^e >$  25 GeV/c\\
         & pseudo-rapidity & $|\eta^e| < $ 2.5\\
         \hline
         \multirow{2}{*}{muons} & Transverse momentum & $p_T^{\mu} >$  25 GeV/c\\
         & Muon pseudo-rapidity & $|\eta^{\mu}| < $ 2.5\\
         \hline
         \multirow{3}{*}{Jets (anti-$k_T$)} &  Radius parameter & $r$ = 0.4\\
         &  Jet transverse momentum & $p_T^j >$ 30 GeV/c\\
         & Jet rapidity & $|\eta^j| < $ 3.0\\
         \hline
          \multicolumn{2}{|c|}{$\Delta\phi$ separation (Leading jet and the $Z$ boson)} & $|{\Delta \phi}_{(Z, jet 1)} |>$ 2.94 radian \\
          \hline
          \multicolumn{2}{|c|}{Second jet veto} & ${p_{T}^{jet 2}}/{p_{T}^{Z}} <$ 15\%\\
         \hline
          \multicolumn{2}{|c|}{Jet-lepton separation} & $\Delta R_{\ell, jet 1} > 0.4$\\
         \hline
    \end{tabular}
    \caption{Selection criteria used in the analysis.}
    \label{ Cuts_table}
\end{table}

The $Z$+ 1-jet events were simulated with Pythia8, Herwig7 and Sherpa2 using merged leading-order (LO) matrix elements supplemented with the PS emissions in the soft-collinear region in all three event generators. The ME along with the PS corrections are implemented for precise event simulation. The reconstruction of the $Z$ boson from the leptonic decay channels were done by selecting the highest $p_T$ lepton pairs in an event. Only events with atleast two oppositely charged leptons present in the final state, satisfying the selection criteria mentioned in Table \ref{ Cuts_table} were selected otherwise the event was vetoed. By using the four momenta information ($p_x$, $p_y$, $p_z$, and $E$) of the lepton anti-lepton pairs, we perform the addition of their four momenta to reconstruct the kinematic properties of the di-lepton pairs.

\section{Results and Discussions}
\label{results}

In this section, simulation results are reported with description. The comparison of the kinematic distributions of the $Z$ boson and jet by applying the selection criteria described in section \ref{event_select}, using three different LO event generators such as Pythia8, Herwig7, and Sherpa2 are reported. Both the electron and muon channel results are shown.


\subsection{Comparison of Kinematic Distributions}
\subsubsection{\texorpdfstring{Lepton and \textit{Z} Boson $p_T$ Distributions}{Transverse moemntum of the Z boson and the lepton candidate}}
\begin{figure}[H]
    \centering
\includegraphics[width= .49\linewidth, height= 7.4cm]{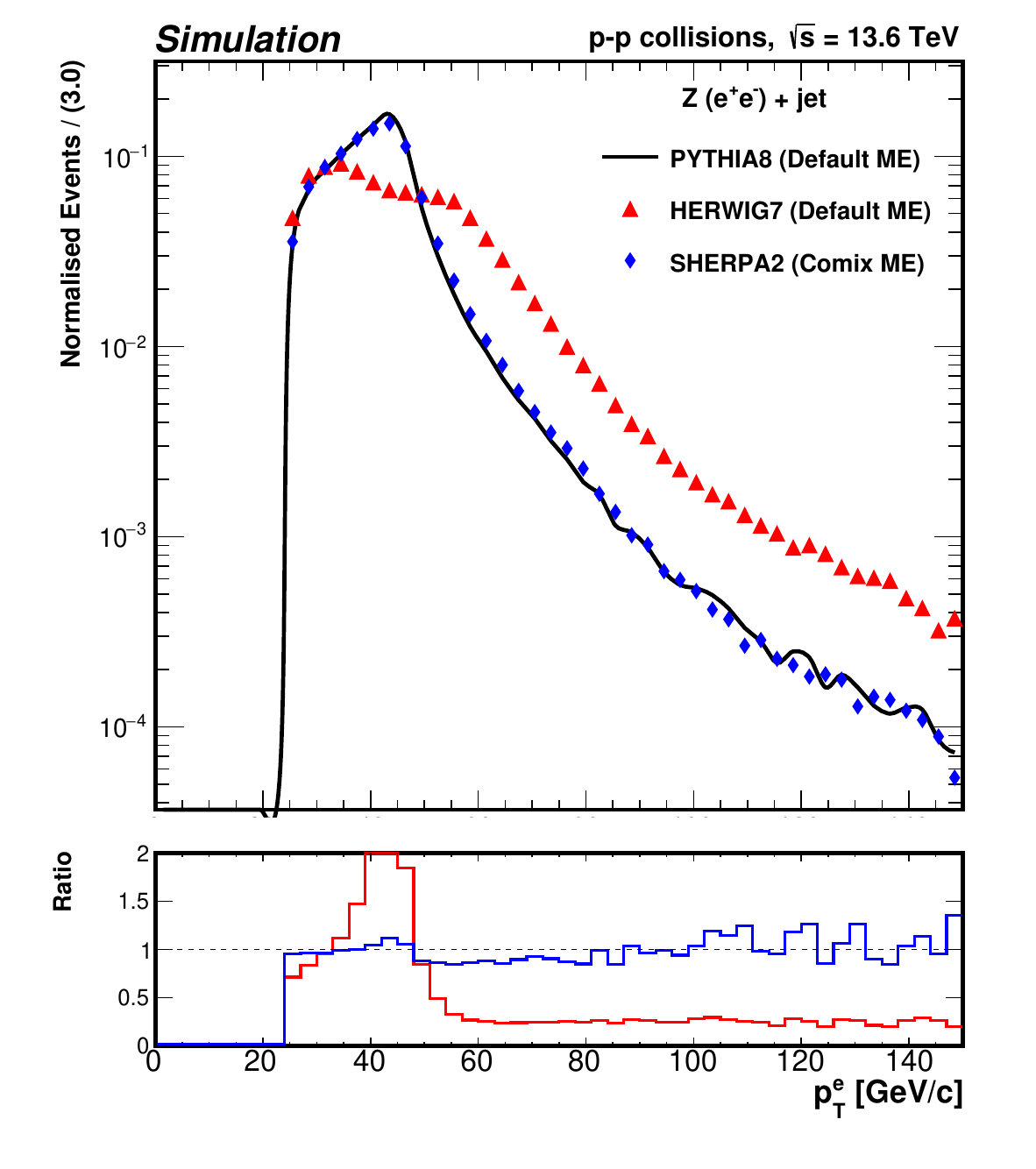}
\includegraphics[width= .49\linewidth, height= 7.4cm]{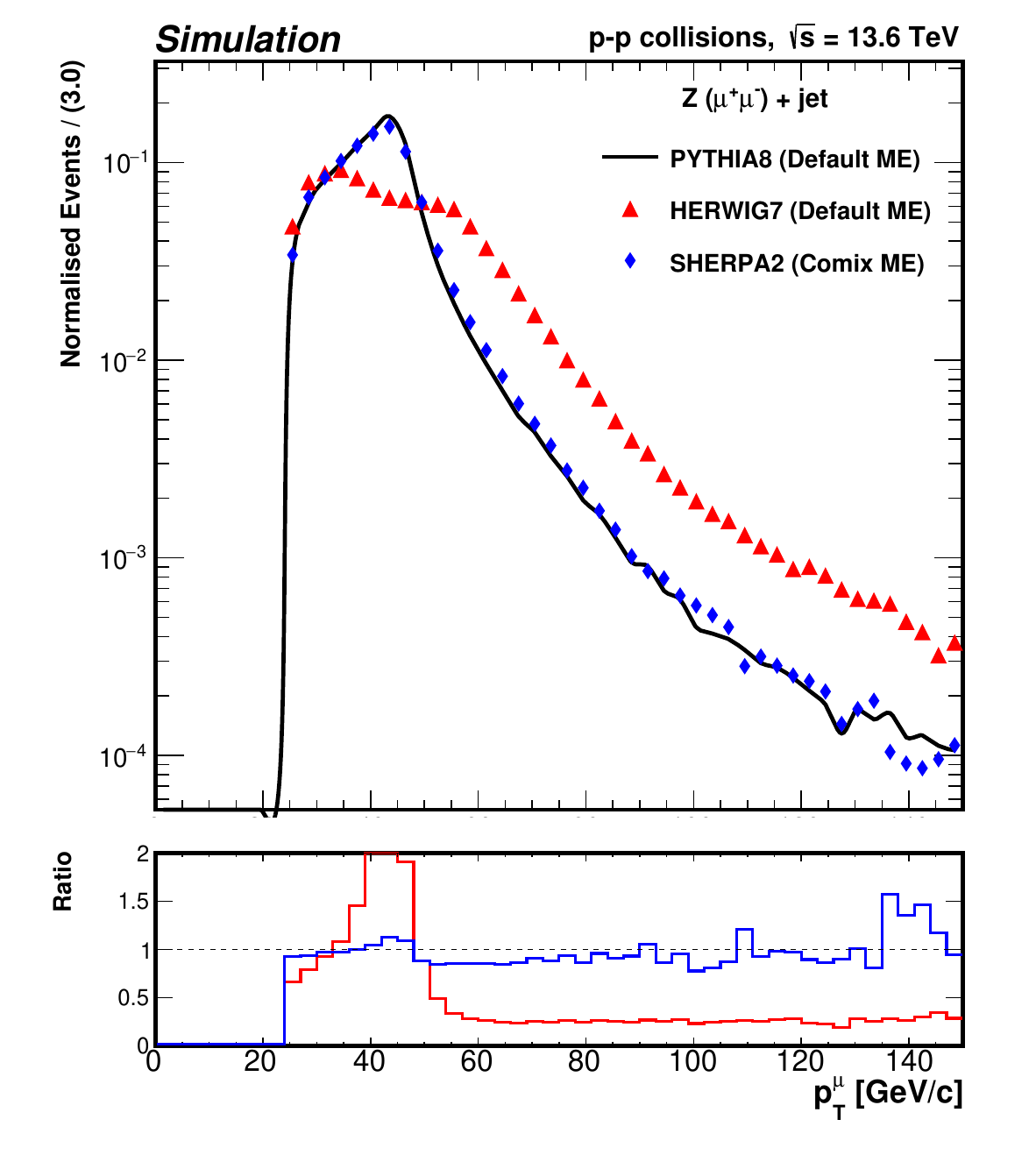}
 \caption{Normalised $p_{T}$ distributions of the lepton candidates in the $Z$+ 1-jet events in electron channel (left) and muon channel (right) using the default ME in both Pythia8 and Herwig7, and COMIX ME in Sherpa2. The lower panel shows the ratio of Pythia8/Herwig7 (red curve) and Pythia8/Sherpa2 (blue curve).}
    \label{pTL_default}
\end{figure}

\begin{figure}[H]
    \centering
\includegraphics[width= .49\linewidth, height= 7.4cm]{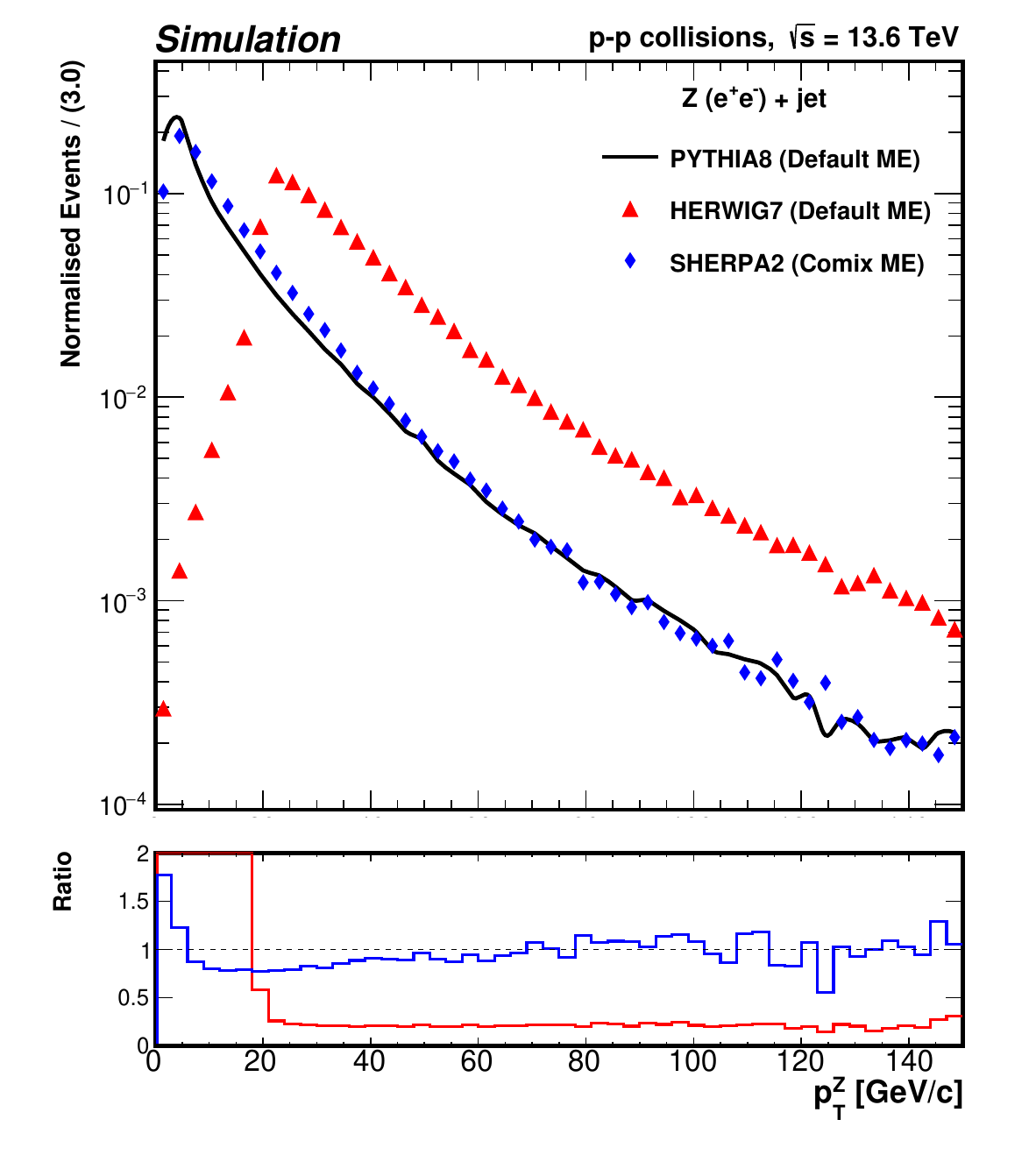}
\includegraphics[width= .49\linewidth, height= 7.4cm]{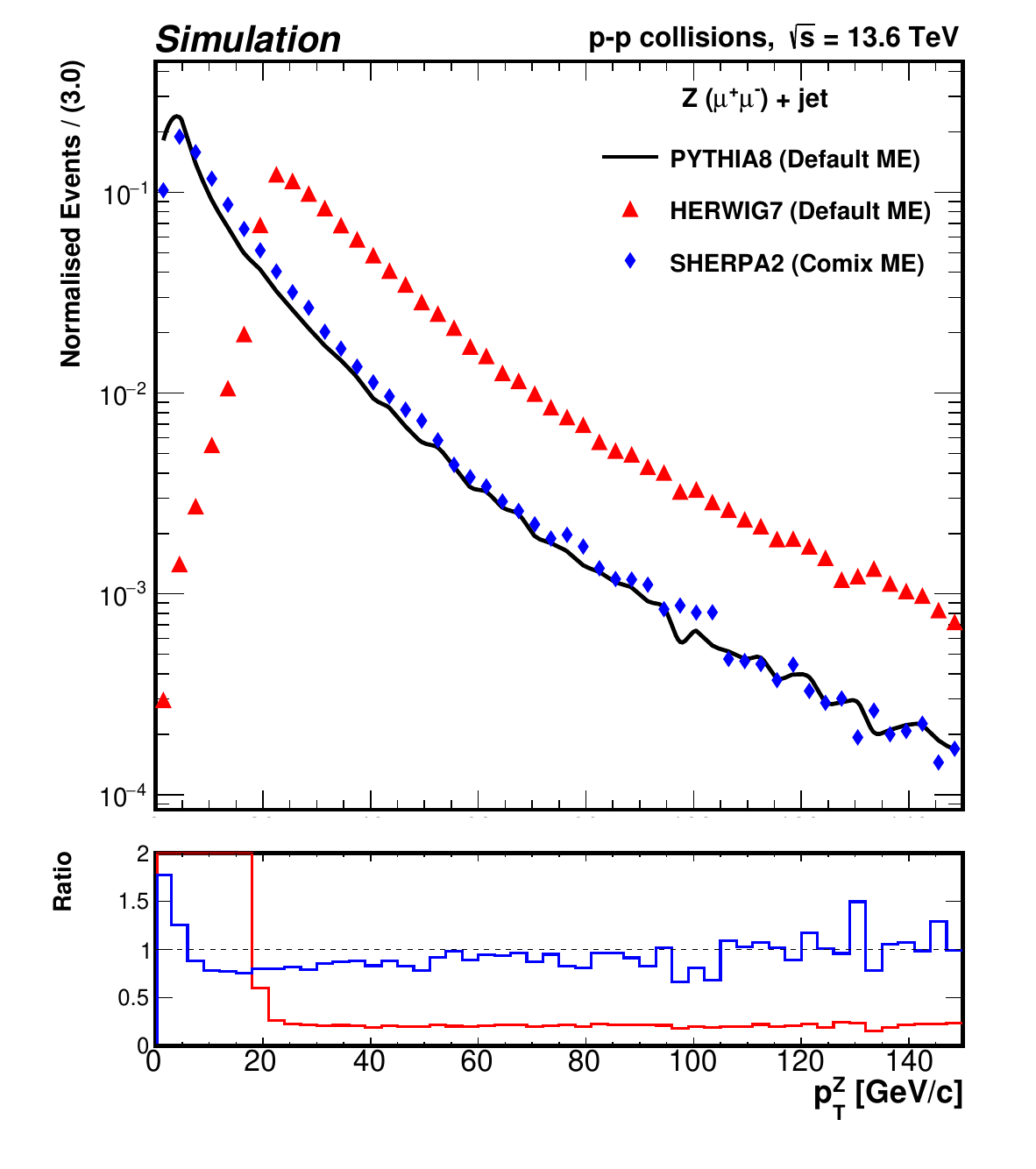}
    \caption{Normalised $p_{T}$ distributions of the $Z$ boson in the $Z$+ 1-jet events in electron channel (left) and muon channel (right) using the Default ME in both Pythia8 and Herwig7, and Comix ME in Sherpa2. The lower panel shows the ratio of Pythia8/Herwig7 (red curve) and Pythia8/Sherpa2 (blue curve).}
    \label{pT_Z_Default}
\end{figure}

Figure \ref{pTL_default} and \ref{pT_Z_Default} displays the comparison between the transverse momentum distributions of the candidate lepton ($p_{T}^{e}$ and $p_{T}^{\mu}$ ) and the $Z$ boson ( $p_{T}^{Z}$) respectively. In case of lepton transverse momentum distributions, we observed a good agreement between the Pythia8 and Sherpa2, however, a significant inconsistency was observed in the Herwig7 predictions, with the default internal ME corrections applied to the PS in all three event generators. Further, the steeply falling pattern of the $p_{T}^{Z}$ spectra in both the electron and muon channels are observed to be better modelled by the Pythia8 and Sherpa2 generators. In the case of Herwig7, the default ME for $Z$ + 1-jet production (MEZJet) appears to have limitations in precisely describing the kinematic properties within the jet-sensitive $Z$+jet process.

\begin{figure}[H]
    \centering
\includegraphics[width= .49\linewidth, height= 7.4cm]{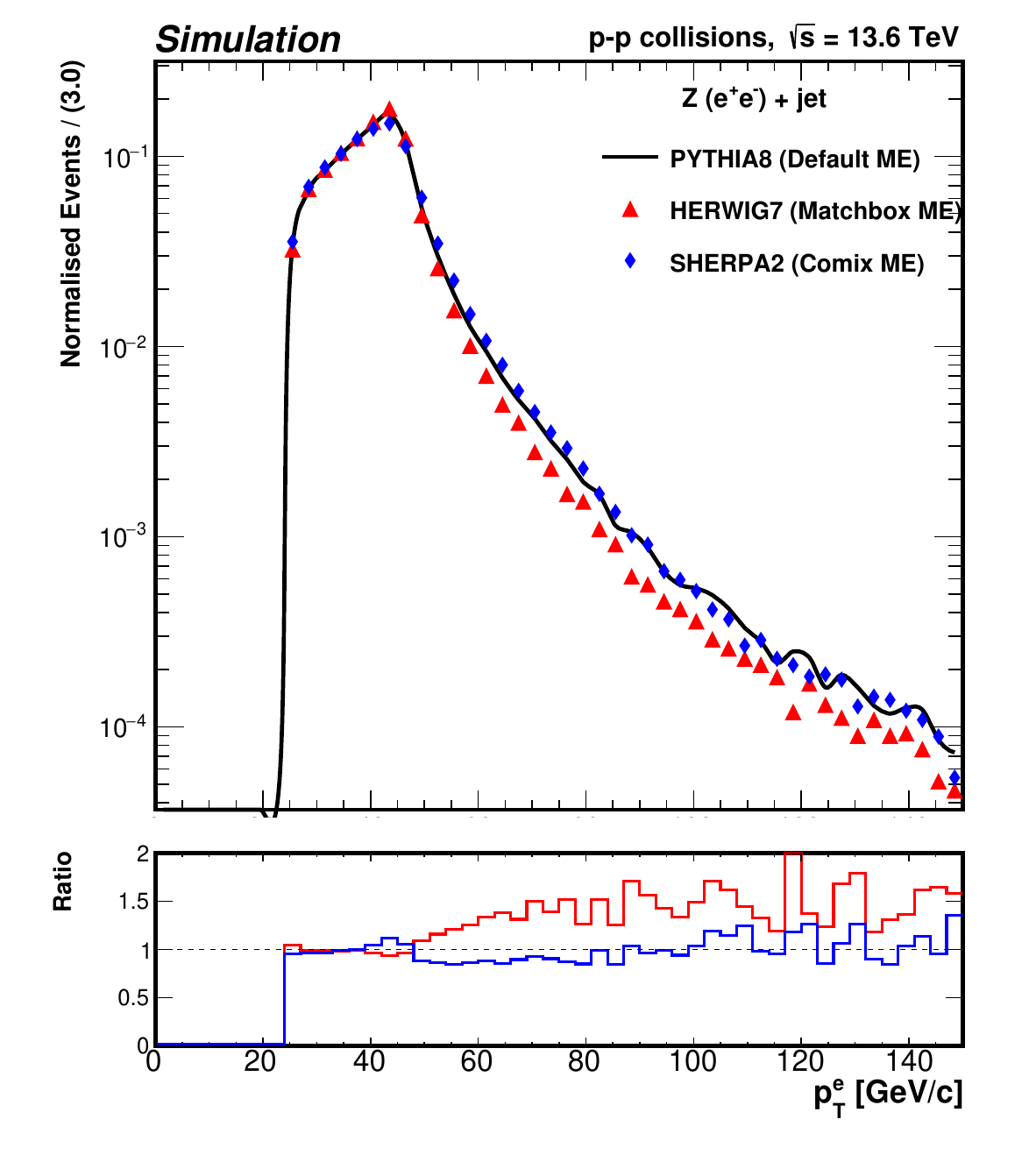}
\includegraphics[width= .49\linewidth, height= 7.4cm]{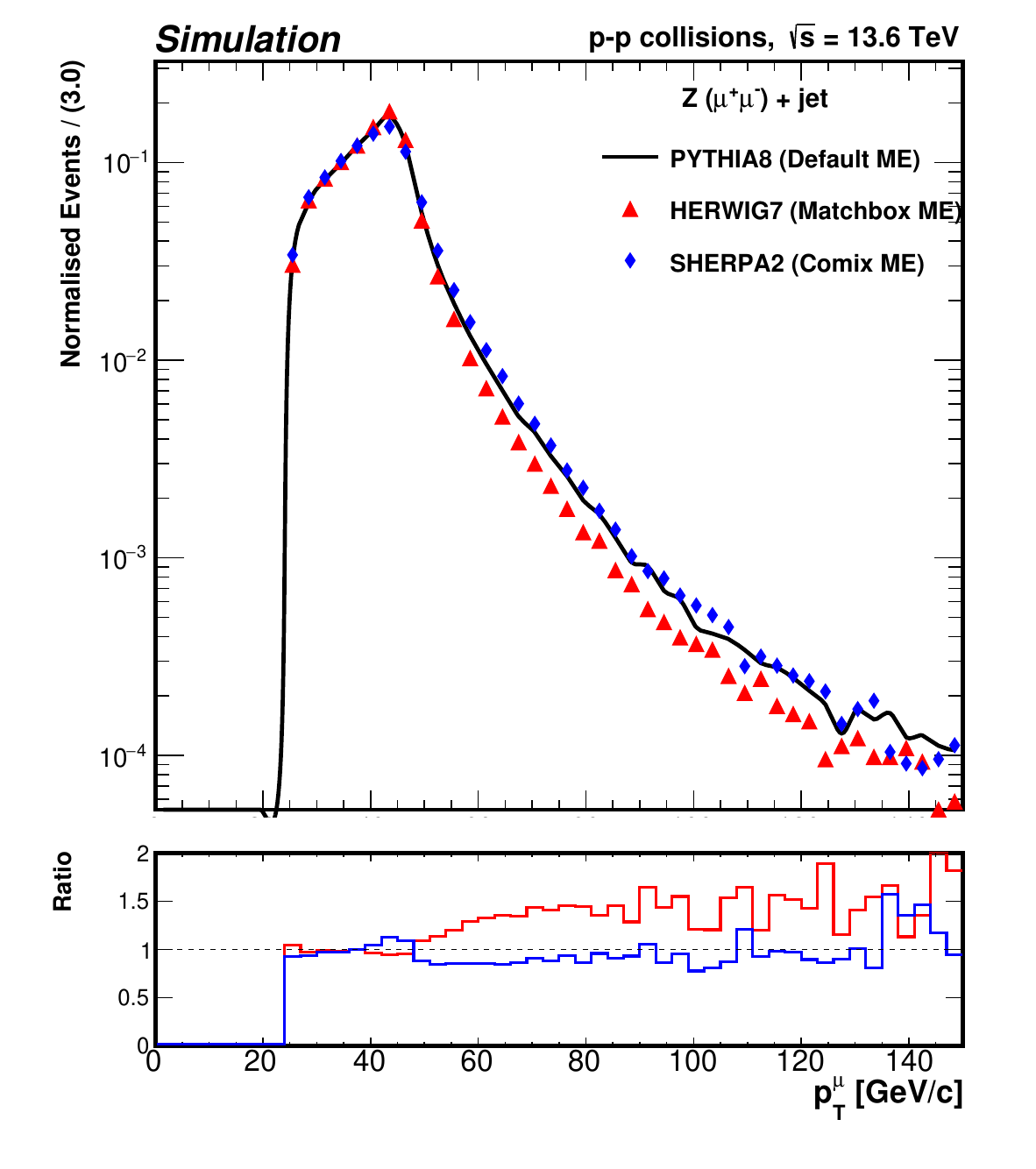}
 \caption{Normalised $p_{T}$ distributions of the lepton candidates in the $Z$+ 1-jet events in electron channel (left) and muon channel (right) using the Default ME in  Pythia8, Matchbox ME in Herwig7, and Comix ME in Sherpa2 MC simulations. The lower panel shows the ratio of Pythia8/Herwig7 (red curve) and Pythia8/Sherpa2 (blue curve).}
    \label{pTL}
\end{figure}

\begin{figure}[H]
    \centering
\includegraphics[width= .49\linewidth, height= 7.4cm]{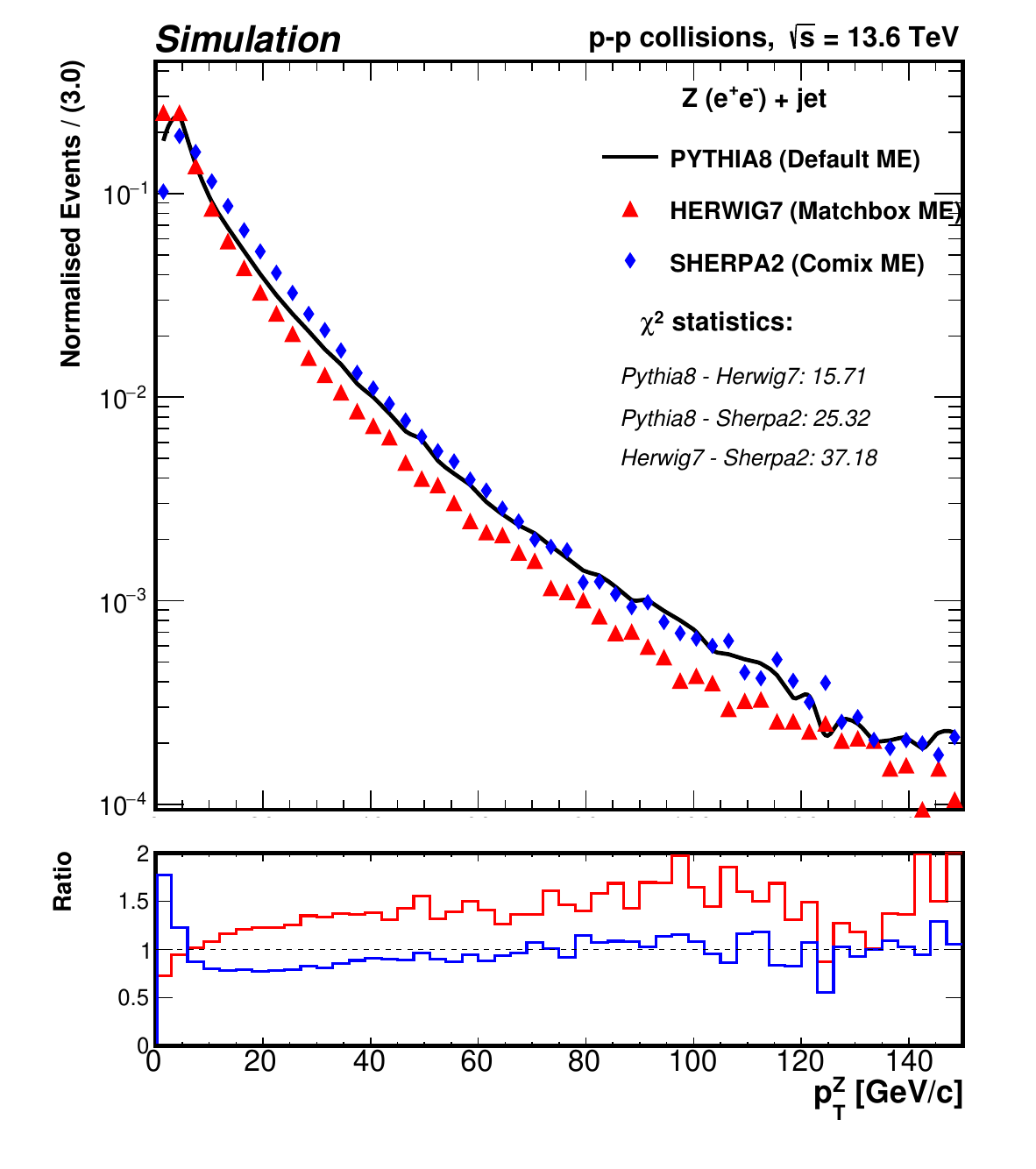}
\includegraphics[width= .49\linewidth, height= 7.4cm]{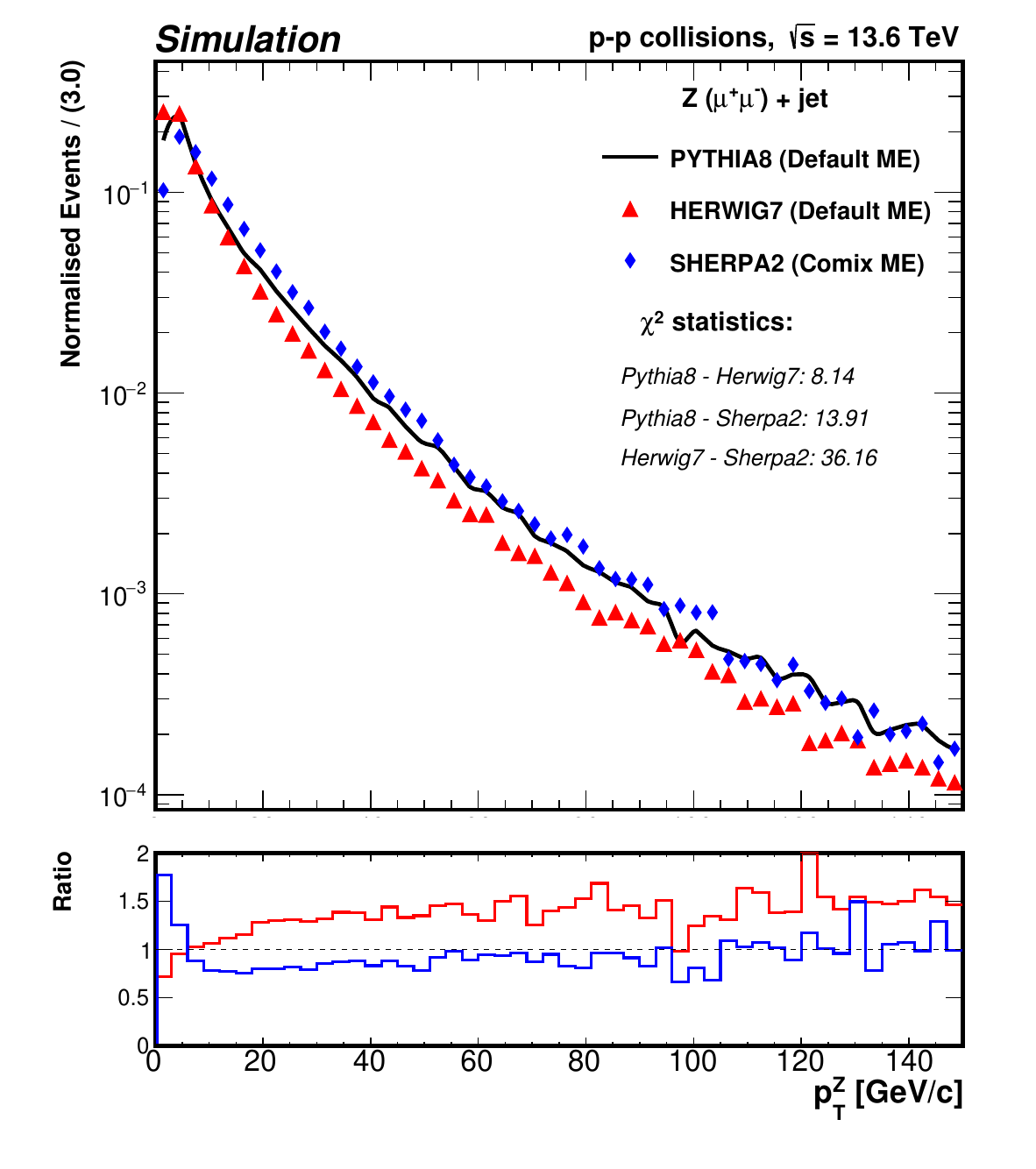}
    \caption{Normalised $p_{T}$ distributions of the $Z$ boson in the $Z$+ 1-jet events with electron channel (left) and muon channel (right) using the Default ME in Pythia8, Matchbox ME in Herwig7, and Comix ME in Sherpa2 MC simulations . The lower panel shows the ratio of Pythia8/Herwig7 (red curve) and Pythia8/Sherpa2 (blue curve).}

    \label{Z_pT}
\end{figure}

In order to understand the cause of the discrepancy, we employed the Matchbox module \cite{Bellm:2015jjp} in Herwig7, which is based on an extended version of ThePEG \cite{Bahr:2008pv,Bertini:2000uh}. This module can perform hard process generation at the level of NLO QCD accuracy. There are pre-encoded ME calculations in the Matchbox module, available for few processes including inclusive Drell-Yan $Z$ and $W$ production, as well as $Z$ and $W$ plus one jet production, that are used in this analysis for LO event generation with the default (`$q$-tilde') shower. We used the default internal ME in Pythia8 with the `Simple' shower \cite{Bierlich:2022pfr} for the parton showering. We compared these MC samples with the Comix ME \cite{Gleisberg:2008fv} in Sherpa2 at LO (with CS Shower \cite{Schumann:2007mg, Sherpa:2019gpd}) with the default ME corrections as implemented in these event generators. Figures \ref{pTL} and \ref{Z_pT} show the $p_T$ distributions of the lepton candidates ($p_{T}^{e}$ and $p_{T}^{\mu}$) and the $Z$ boson reconstructed from the electron channel and the muon channel in the $Z$+ 1-jet process using these configurations.

Now, from the Figures \ref{pTL} and \ref{Z_pT} we note that, although the three event generators have different implementations in their modelling and showering, they all effectively produce the $p_T$ distributions of the leptons and $Z$ boson similarly in both decay channels. The overall shape of the kinemtic distributions improved with the Matchbox ME in Herwig7 with a better agreement was observed with Pythia8 and Sherpa2 predictions over the whole $p_{T}$ range.

\subsubsection{\textit{Z} Boson Mass Distributions}
\begin{figure}
    \centering
\includegraphics[width= .49\linewidth, height= 7.4cm]{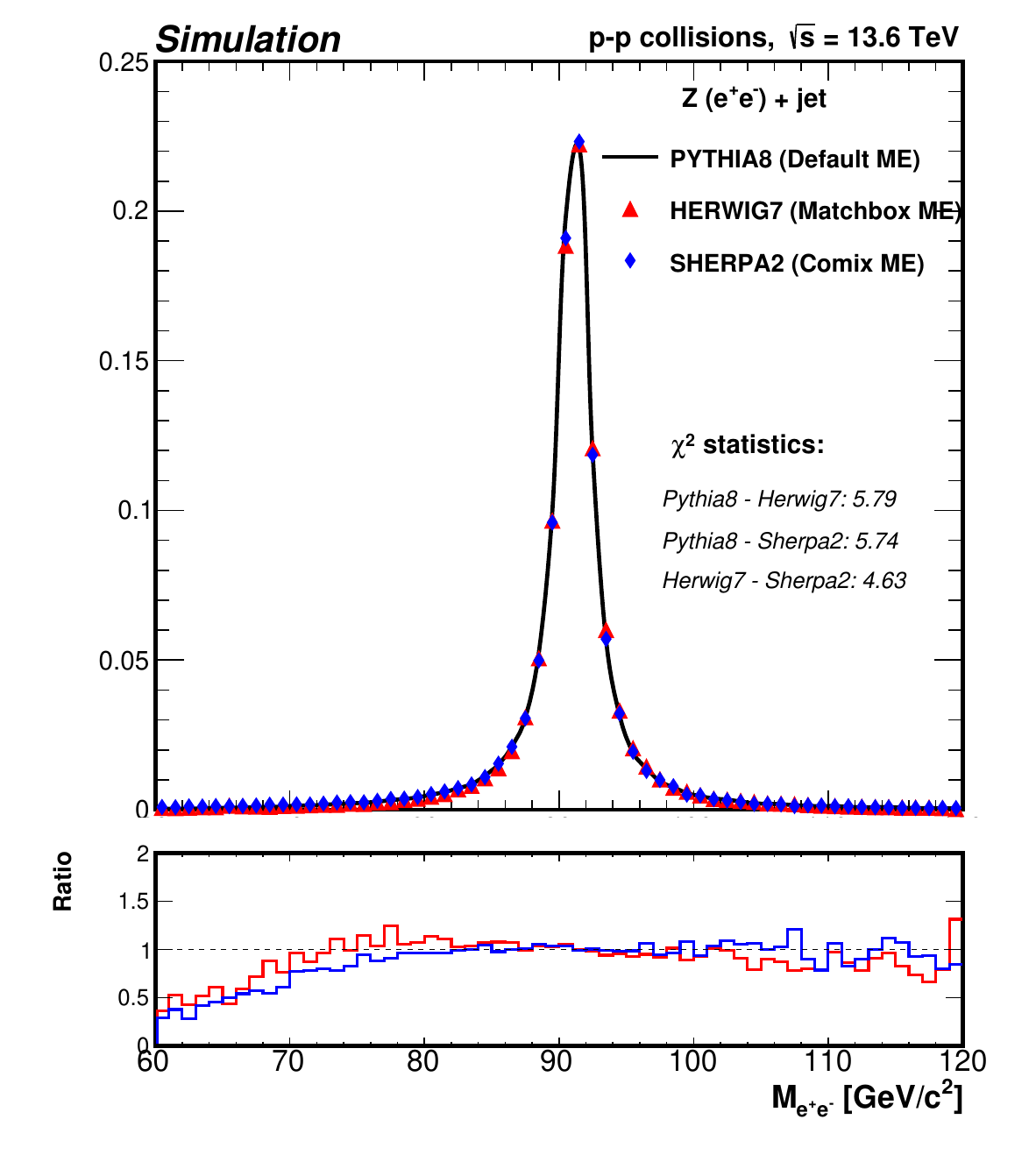}
\includegraphics[width= .49\linewidth, height= 7.4cm]{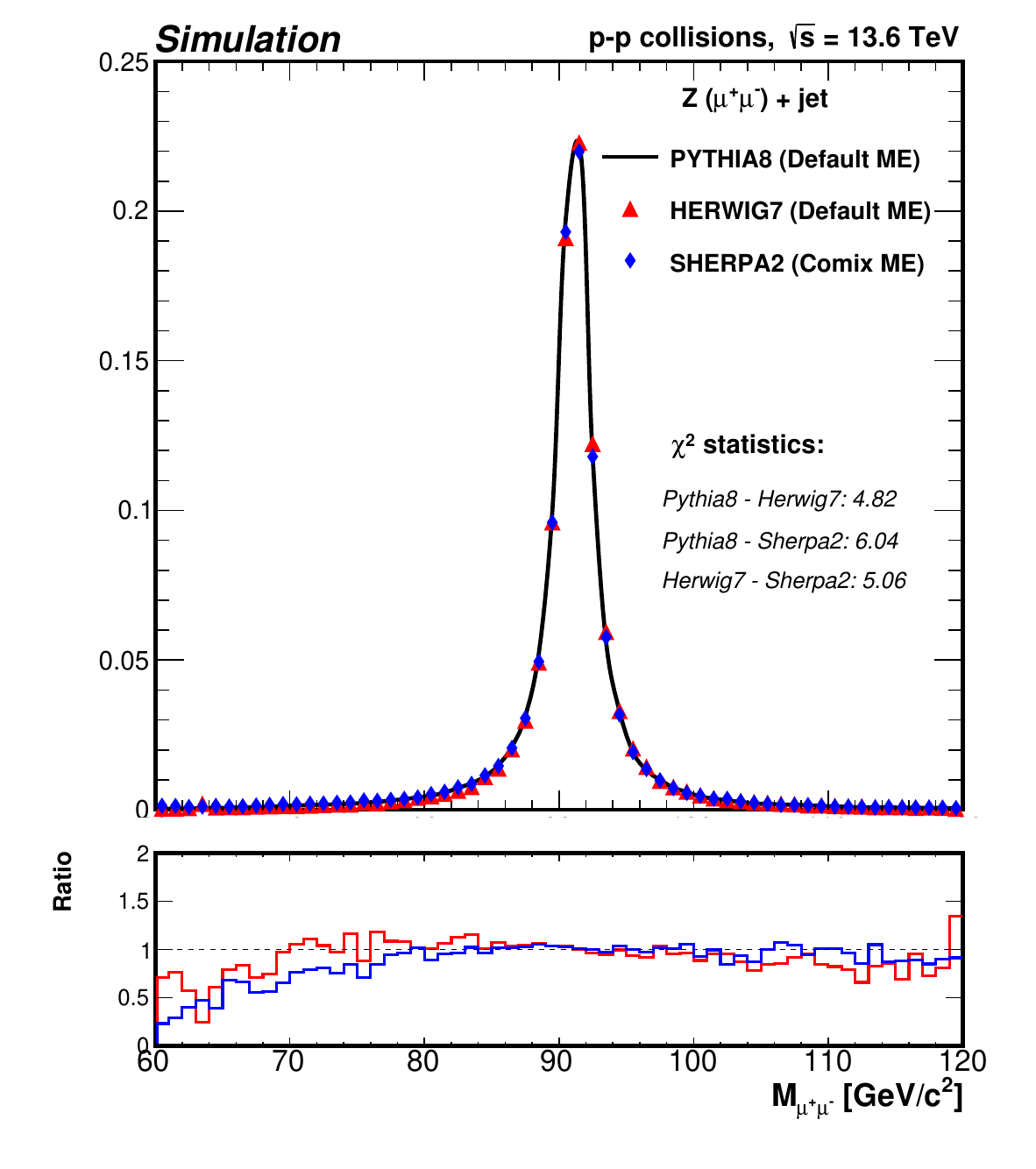}
    \caption {Normalised  di-lepton invariant mass, $M_{\ell^{+} \ell^{-}}$ ($Z$ boson mass) distribution in the electron channel (left) and muon channel (right) with the Default ME in Pythia8, Matchbox ME in Herwig7, and Comix ME in Sherpa2 MC simulations. The lower panels in the plots show the ratio of Pythia8/Herwig7 (red curve) and Pythia8/Sherpa2 (blue curve).}
    \label{Z_mass}
\end{figure}

\begin{figure}
    \centering
\includegraphics[width= .49\linewidth, height= 7.4cm]{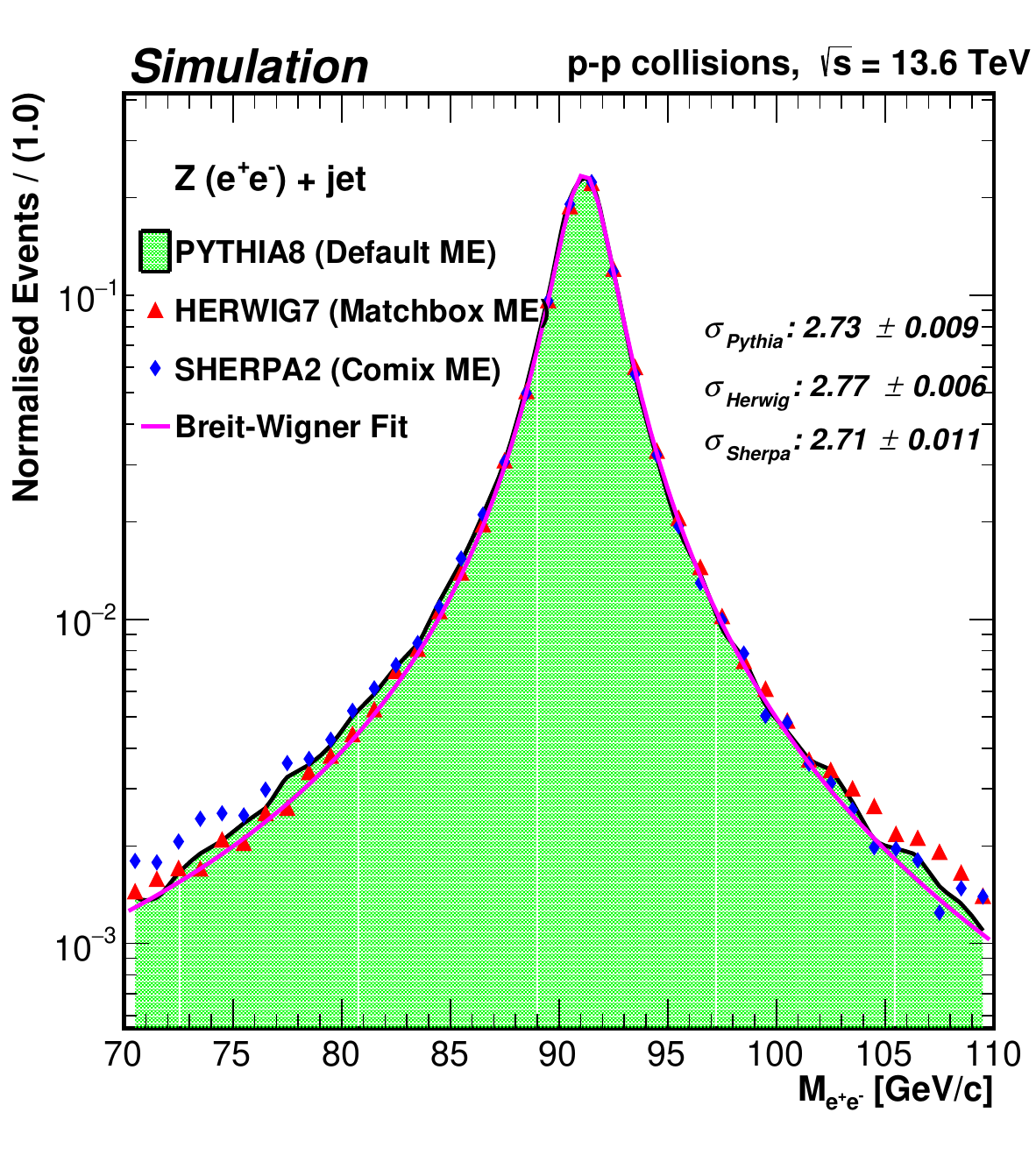}
\includegraphics[width= .49\linewidth, height= 7.4cm]{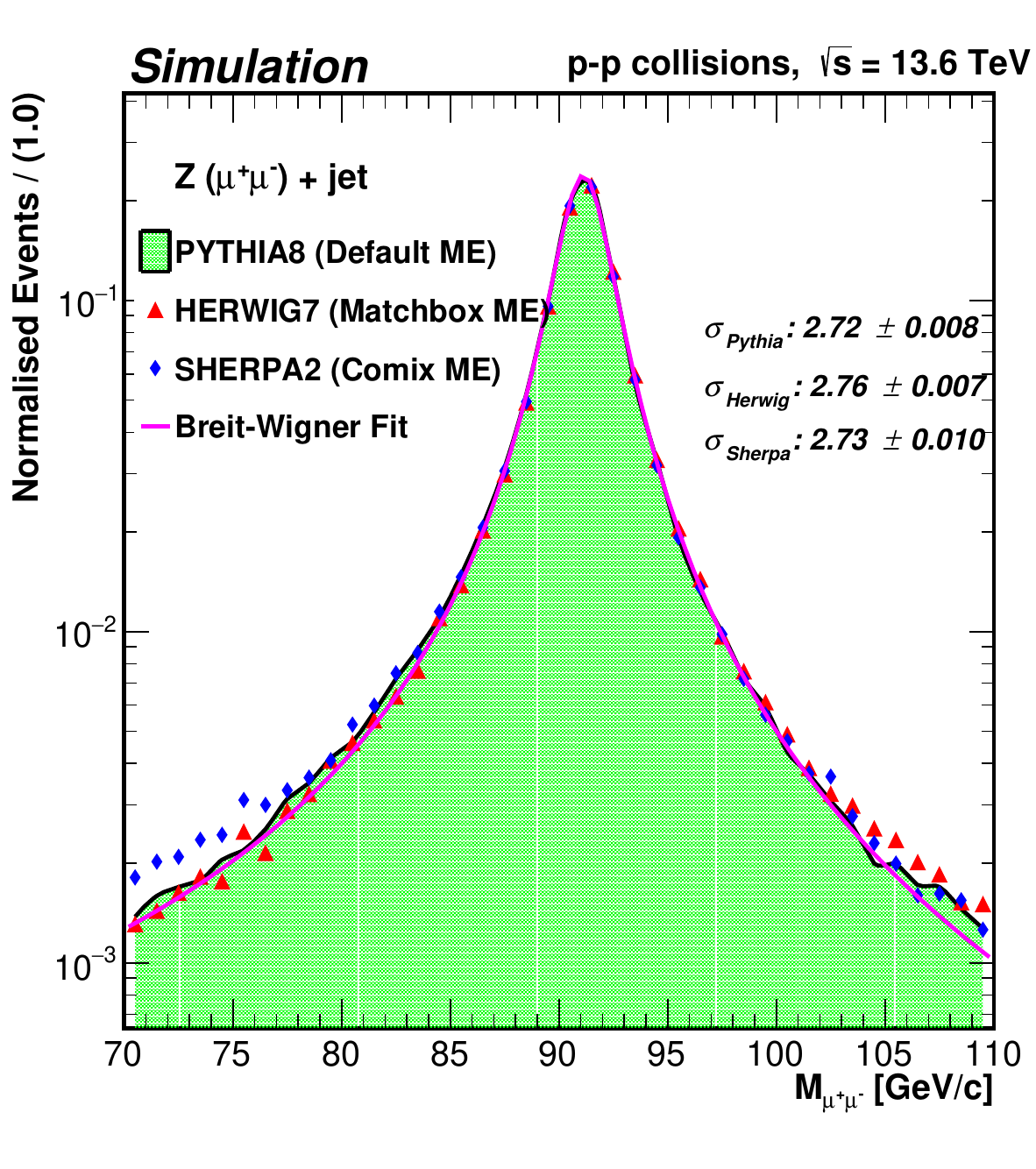}
    \caption{Di-lepton invariant mass, $M_{\ell^{+} \ell^{-}}$ distributions in the electron channel (left) and muon channel (right) fitted with the Breit-Wigner function in the $M_{\ell^{+} \ell^{-}}$ mass range of 70$-$110 GeV/$c^2$ in different MC simulations.}
    \label{mZ_BW}
\end{figure}

A study of the \textit{Z} boosn mass reconstruction by taking di-lepton invariant mass has been conducted in both the electron and muon decay channels, following the lepton selections outlined in Section \ref{event_select}. The di-lepton invariant mass distributions in both decay channels are similarly represented in all three event generators, with the invariant mass peaking near the $Z$ boson mass in all three MC simulations, as illustrated in Figure \ref{Z_mass}. Nonetheless, a discrepancy was observed in the lower mass region of the $Z$ boson in both decay channels. Further, the di-lepton invariant mass distributions were fitted with the Breit-Weigner function\cite{BW} over the $M_{\ell^{+} \ell^{-}}$ mass range of 70$-$110 GeV/$c^2$, as shown in Figure \ref{mZ_BW}. The Gaussian width ($\sigma$) of the fitted function for different MC samples for electron and muon decay channels are shown in Table \ref{sigma_mZ}. The fitted $\sigma$ values show significant consistency across all the MC predictions as well as in both decay channels. A more comprehensive quantitative analysis of the kinematic distributions of the $Z$ boson is discussed in Section \ref{stats}.

\begin{table}[h!]
\centering
\begin{tabular}{|c|c|c|c|}
\hline
\textbf{Gaussian width $\sigma$ (in GeV)} & \textbf{Pythia8} & \textbf{Herwig7}&\textbf{Sherpa2}\\
\hline
electron channel  & 2.73$\pm$0.009 & 2.77$\pm$0.006 & 2.71$\pm$0.011 \\
\hline
muon channel & 2.72$\pm$0.008 & 2.76$\pm$0.007 & 2.73$\pm$0.010 \\
\hline
\end{tabular}
\caption{Gaussian width ($\sigma$) (in GeV) of reconstructed di-lepton invariant mass distributions fitted with Breit-Wigner function from $Z$+ 1-jet events with the Default ME in Pythia8, Matchbox ME in Herwig7, and Comix ME in Sherpa2 MC simulations.}
\label{sigma_mZ}
\end{table}

 
\newpage

\subsubsection{\texorpdfstring{Jet Multiplicity and Leading Jet $p_{T}$ Distributions}{Transverse Momentum of the Leading Jet}}
{}

\begin{figure}[H]
    \centering
\includegraphics[width= .49\linewidth, height= 7.4cm]{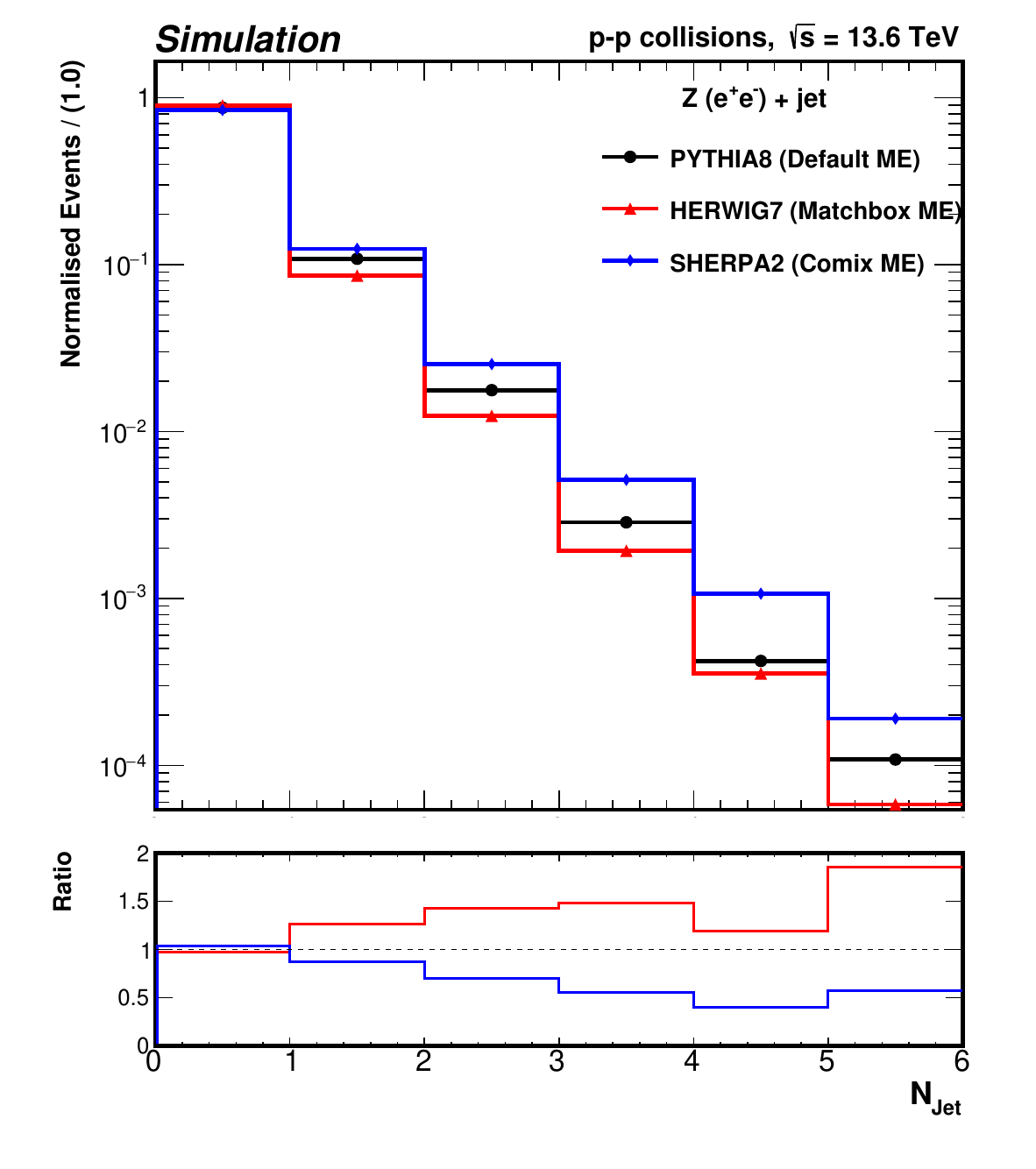}
\includegraphics[width= .49\linewidth, height= 7.4cm]{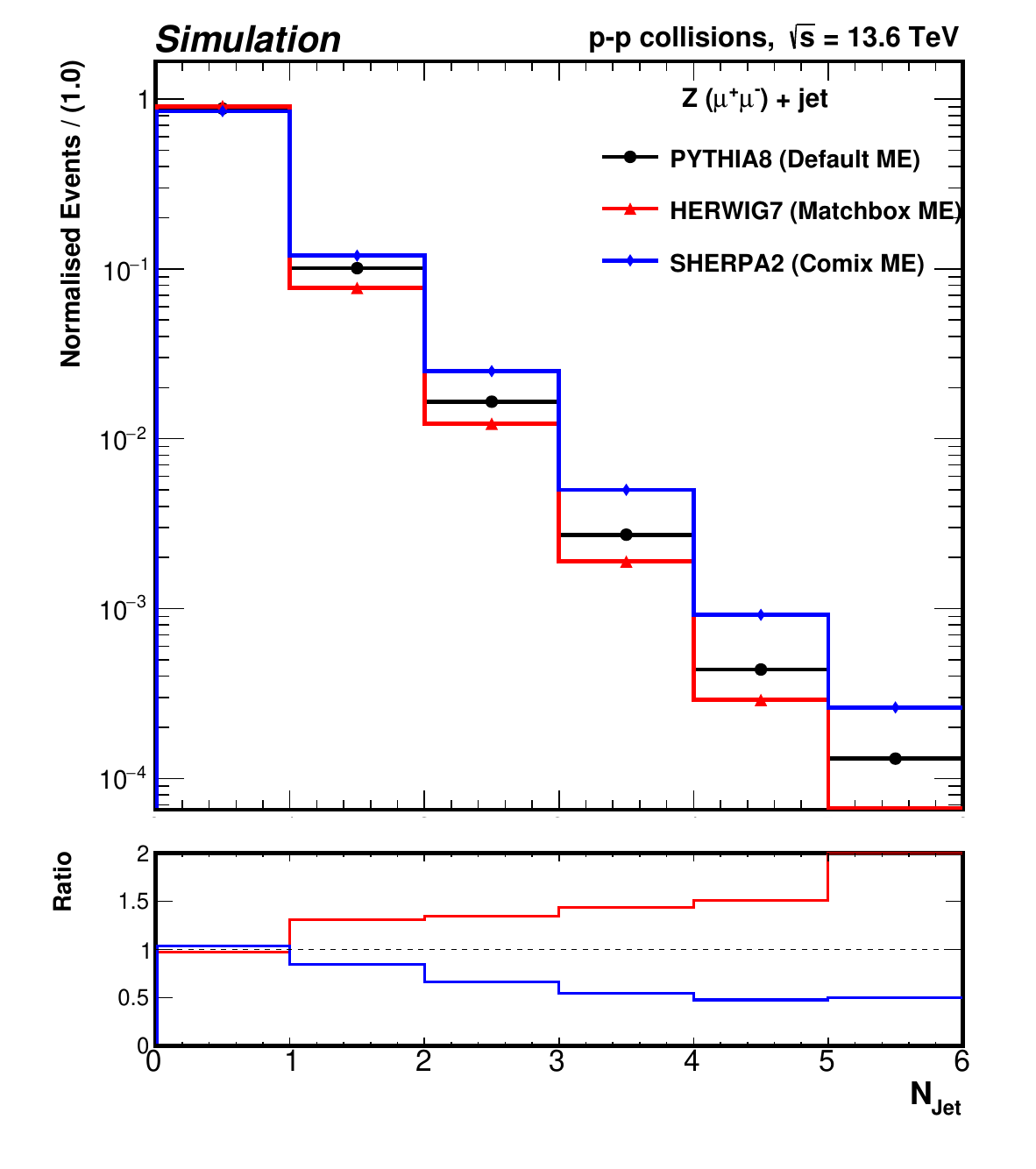}
    \caption{Normalised jet multiplicity distributions with jet $p_T$ greater than 30 GeV/c and within $|\eta| < 3.0$ range in $Z$+ 1-jet events in the electron channel (left) and muon channel (right) of the $Z$ boson decay with the Default ME in Pythia8, Matchbox ME in Herwig7, and Comix ME in Sherpa2. The lower panel shows the ratio of Pythia8/Herwig7 (red curve) and Pythia8/Sherpa2 (blue curve).}
    \label{NJET}
\end{figure}

\begin{figure}[H]
    \centering
\includegraphics[width= .49\linewidth, height= 7.4cm]{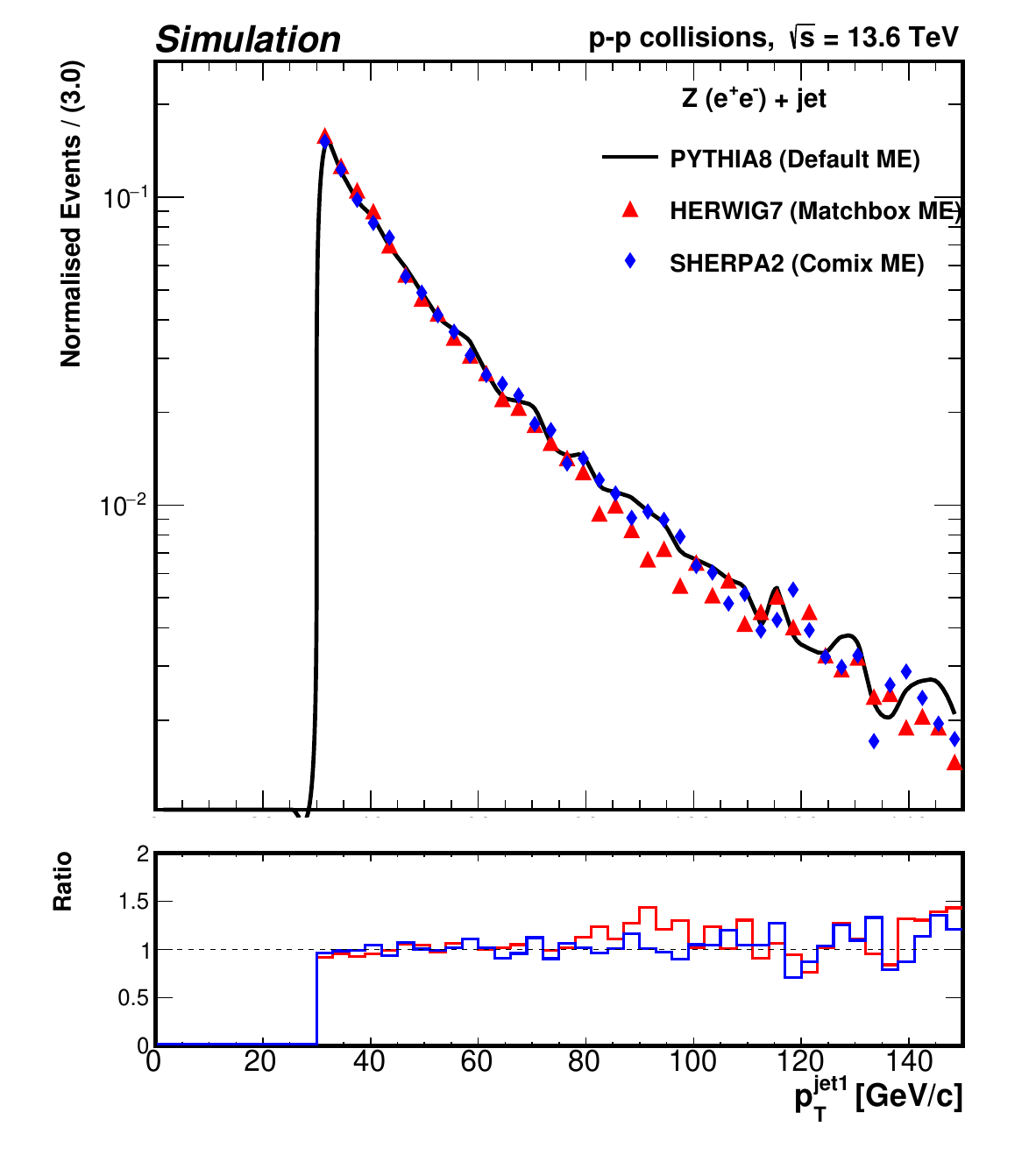}
\includegraphics[width= .49\linewidth, height= 7.4cm]{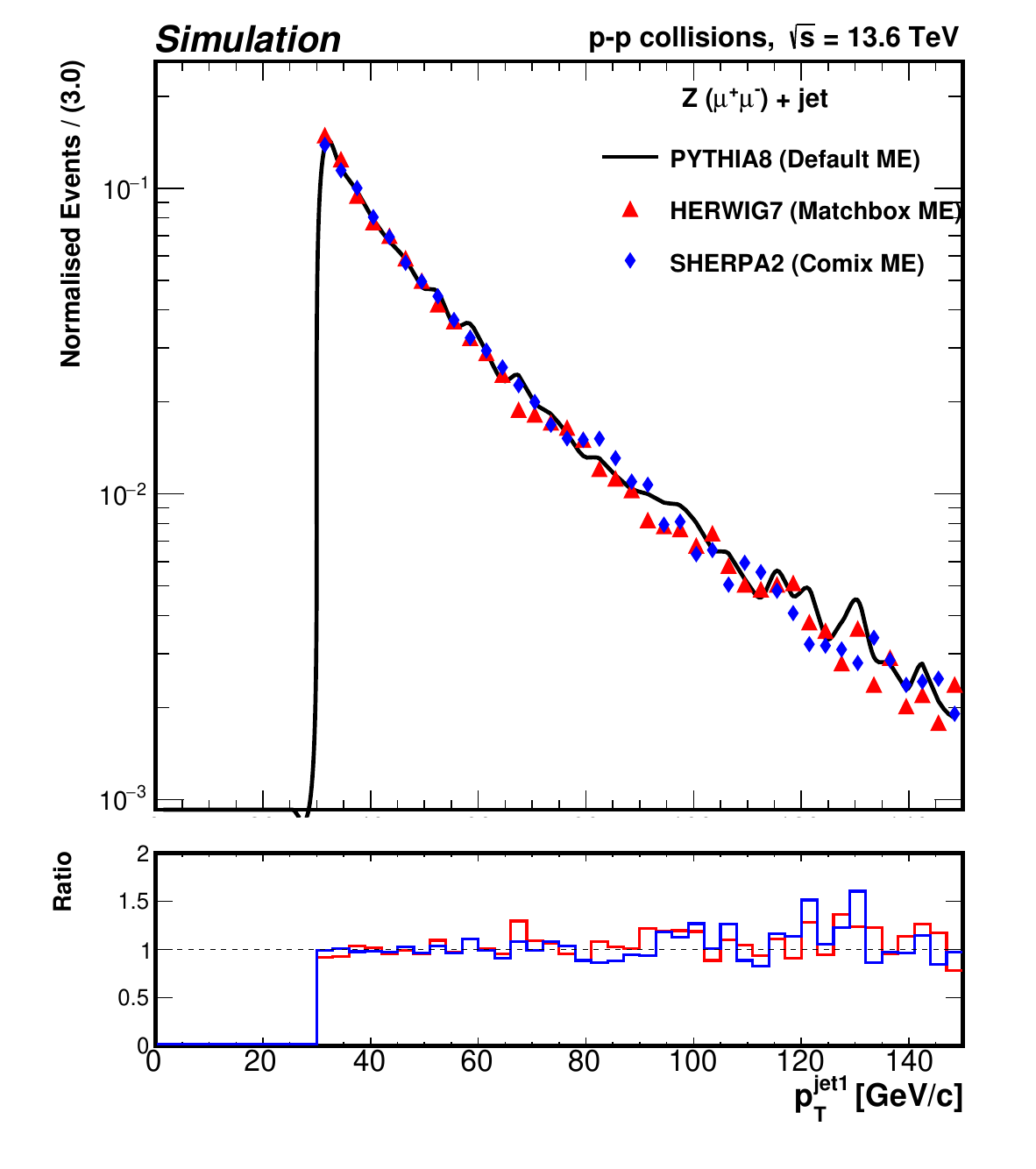}
    \caption{Normalised transverse momentum distributions of the leading jet with $p_T >$ 30 GeV/c and within $|\eta| < 3.0$ range in $Z$+ 1-jet events for electron channel (left) and muon channel (right) of the $Z$ boson decay with using the Default ME in Pythia8, Matchbox ME in Herwig7, and Comix ME in Sherpa2. The lower panel shows the ratio of Pythia8/Herwig7 (red curve) and Pythia8/Sherpa2 (blue curve).}
    \label{pT_J1}
\end{figure}

The normalised jet multiplicity ($N_{Jet}$) distributions with jet $p_T$ threshold of 30 GeV/c are presented in Figure \ref{NJET} for electron and muon channels. The normalised leading jet transverse momentum ($p_{T}^{jet1}$) distributions with leading jet $p_T$ greater than 30 GeV/c are presented in Figure \ref{pT_J1} by taking $Z(e^{+}e^{-}$) + 1-jet and $Z(\mu^{+}\mu^{-}$) + 1-jet events. All three simulations with the Default ME in Pythia8, Matchbox ME in Herwig7, and Comix ME in Sherpa2 seem to produce similar $N_{Jet}$ shapes (upto 2 jets), with Sherpa2 producing a slightly higher number of jets per event compared to the other two MC generators.
The leading jet $p_T$ was found to be very accurately described in all three simulations with the Default ME in Pythia8, Matchbox ME in Herwig7, and Comix ME in Sherpa2. The ratios of Pythia8/Herwig7 and Pythia8/Sherpa2 in case of $p_{T}^{jet1}$ are very close to 1.

\subsubsection{Lepton - Jet separation}
To mitigate the contamination of fake leptons from jets and enhance the purity of leptons for \textit{Z} boson reconstruction, we apply a $|\Delta R| > 0.4$ cut between the leading jet and leptons. 

\begin{figure}[H]
    \centering
\includegraphics[width= .49\linewidth, height= 7.4cm]{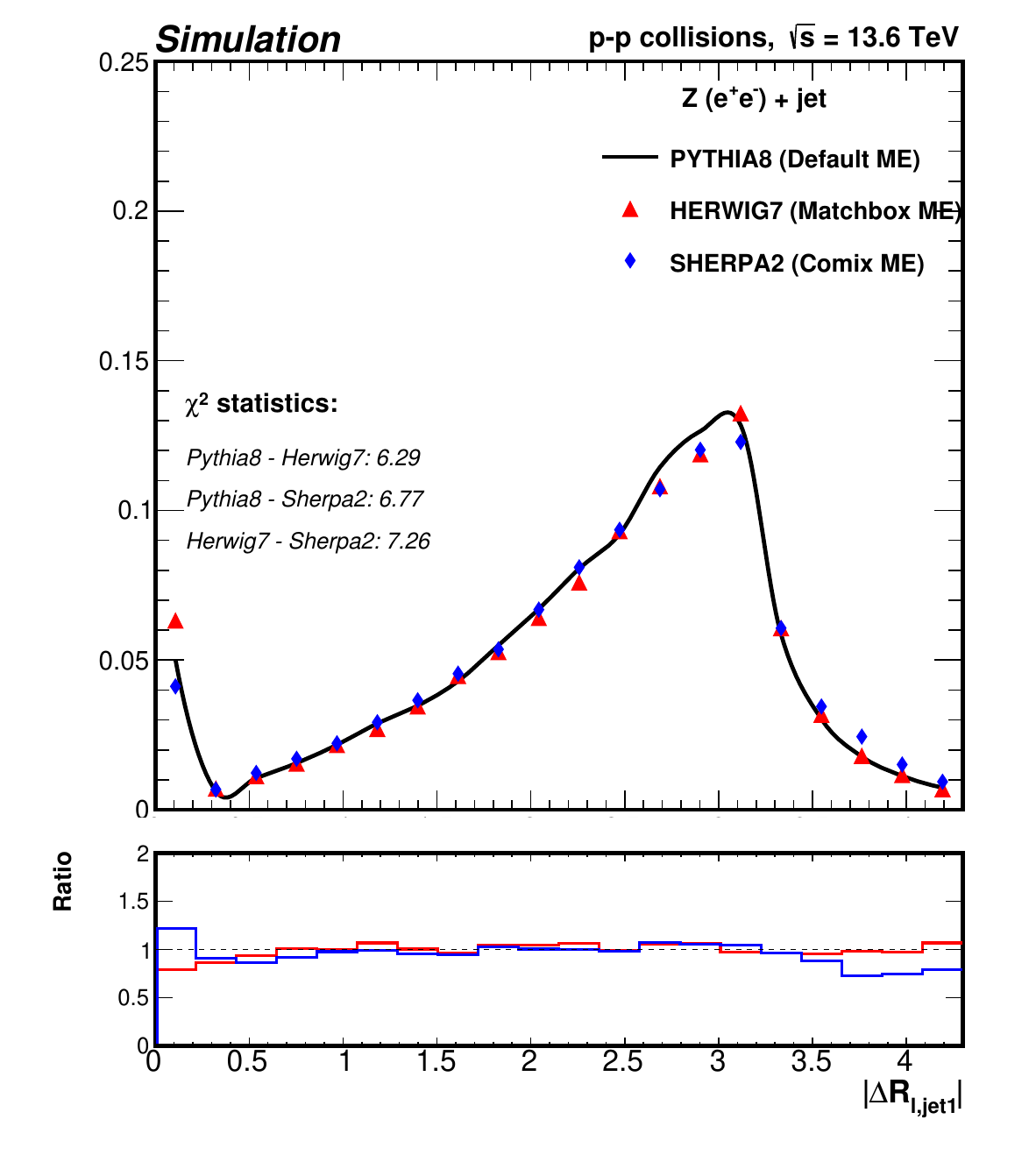}
\includegraphics[width= .49\linewidth, height= 7.4cm]{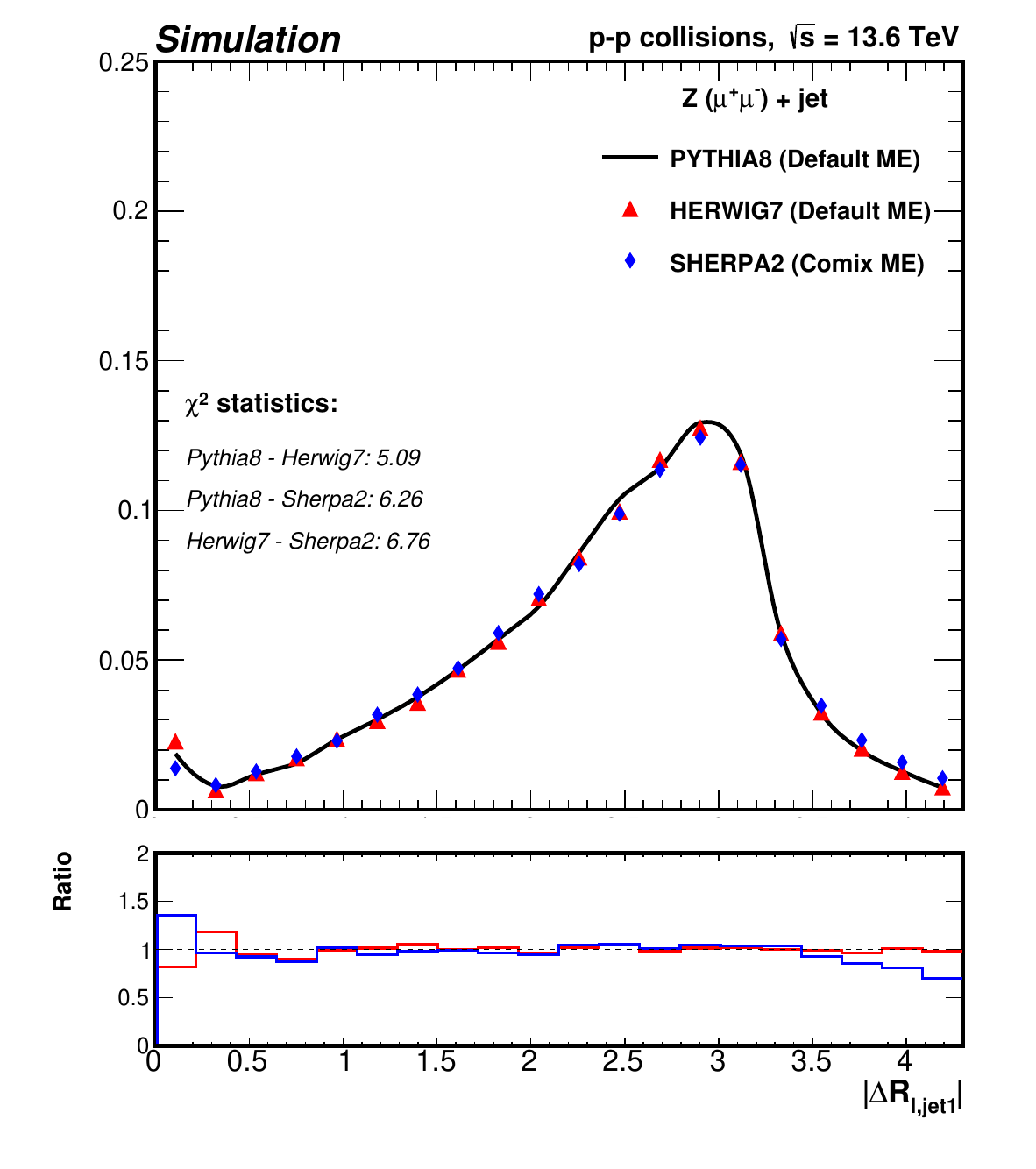}
    \caption{Normalised $|\Delta R|$ separation between the leading jet and the candidate lepton in $Z$+ 1-jet process for electron channel (left) and muon channel (right) of the $Z$ boson decay with Default ME in Pythia8, Matchbox ME in Herwig7, and Comix ME in Sherpa2. The lower panel shows the ratio Pythia8/Herwig7 (red curve) and Pythia8/Sherpa2 (blue curve).}
    \label{delta_R}
\end{figure}

Figure \ref{delta_R} shows the normalised distributions of the distance between the lepton candidates and the leading jet ($\Delta R_{\ell, jet1}$ = $\sqrt{(\Delta \eta_{\ell,jet1})^2 + (\Delta \phi_{\ell,jet1})^2}$) in the pseudo rapidity-azimuthal plane. The $|\Delta R_{\ell, jet1}|$ distribution observed to be similarly modelled with the Default ME in Pythia8, Matchbox ME in Herwig7, and Comix ME in Sherpa2. 


\subsubsection{Back-to-Back Topology of \textit{Z} Boson and the Leading Jet}
\begin{figure}[H]
    \centering
\includegraphics[width= .49\linewidth, height= 7.4cm]{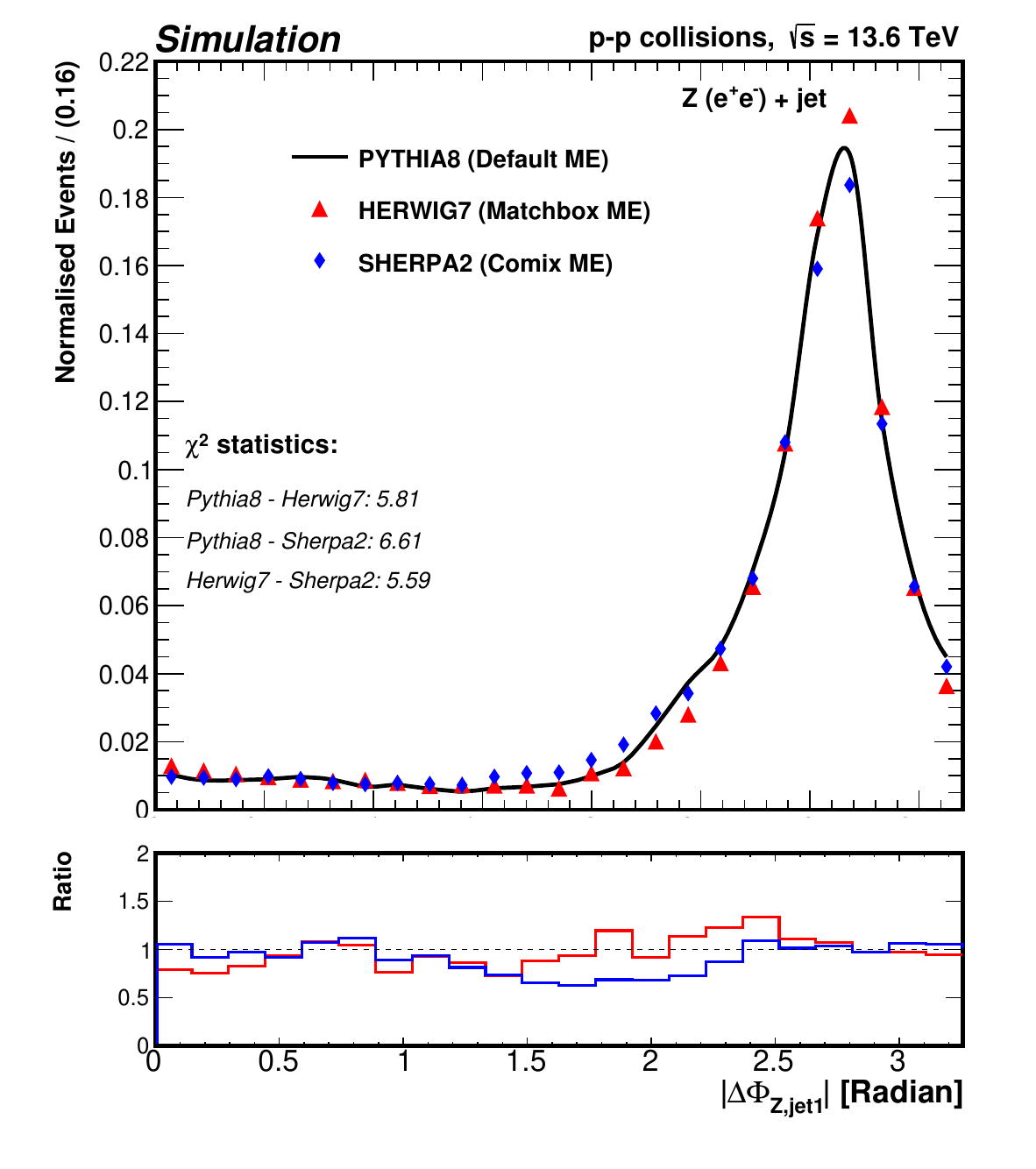}
\includegraphics[width= .49\linewidth, height= 7.4cm]{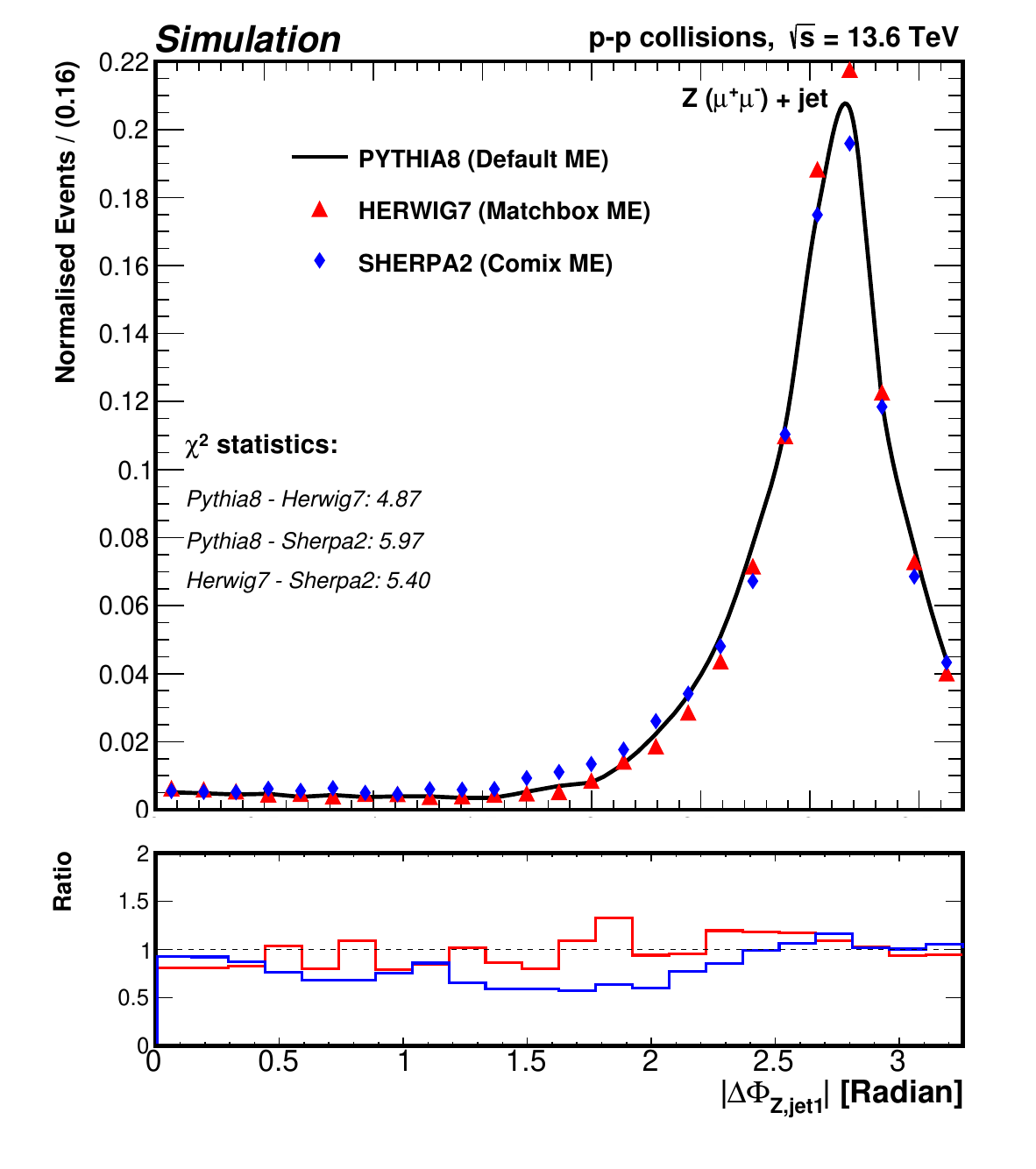}
    \caption{Normalised $|\Delta \phi|$ distribution between the leading jet and $Z$ boson in $Z$+ 1-jet process for electron channel (left) and muon channel (right) of the $Z$ boson decay with the Default ME in Pythia8, Matchbox ME in Herwig7, and Comix ME in Sherpa2. The lower panels show the ratio Pythia8/Herwig7 (red curve) and Pythia8/Sherpa2 (blue curve).}
    \label{delta_phi}
\end{figure}

\begin{figure}[H]
    \centering
\includegraphics[width= .49\linewidth, height= 7.4cm]{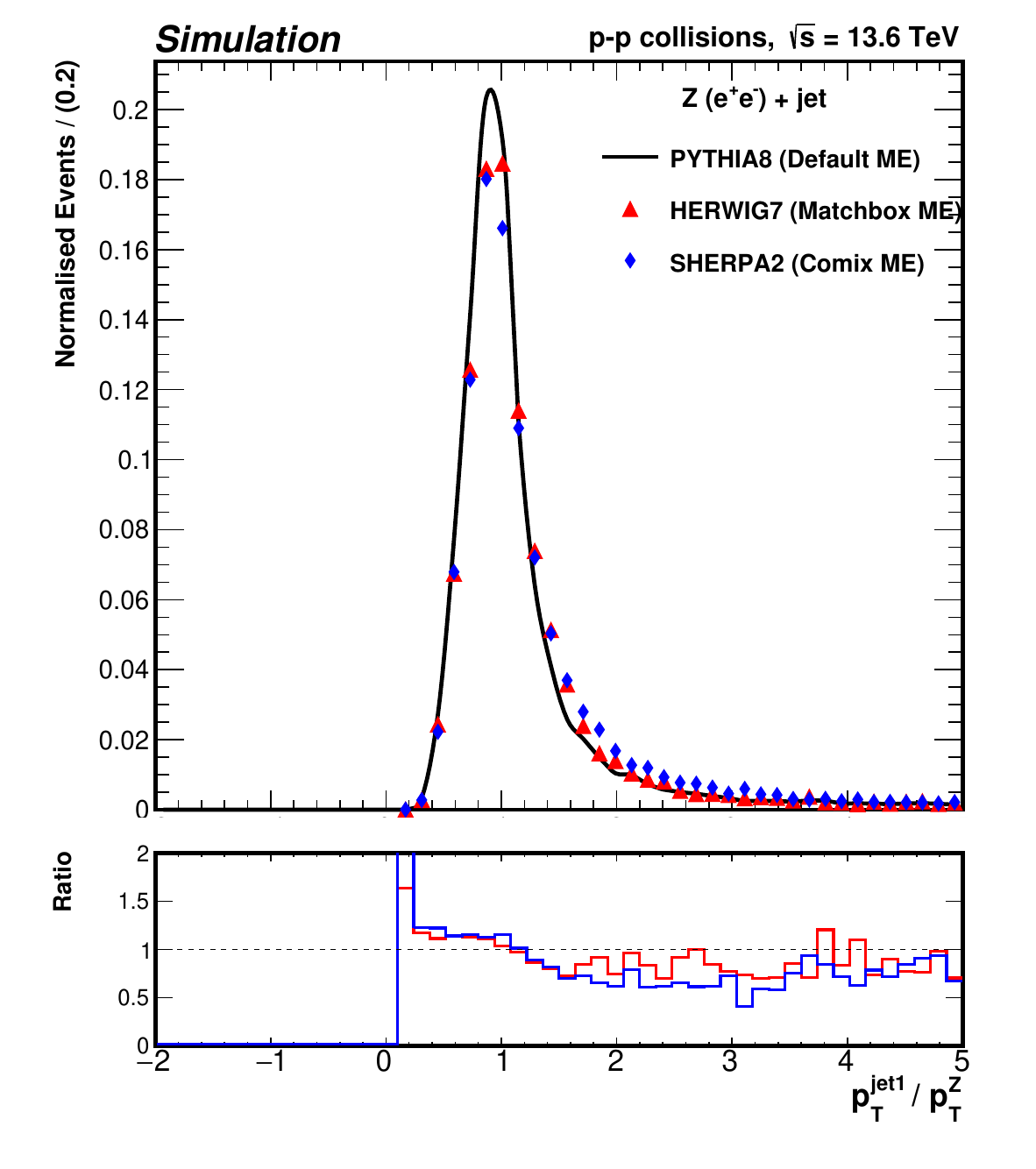}
\includegraphics[width= .49\linewidth, height= 7.4cm]{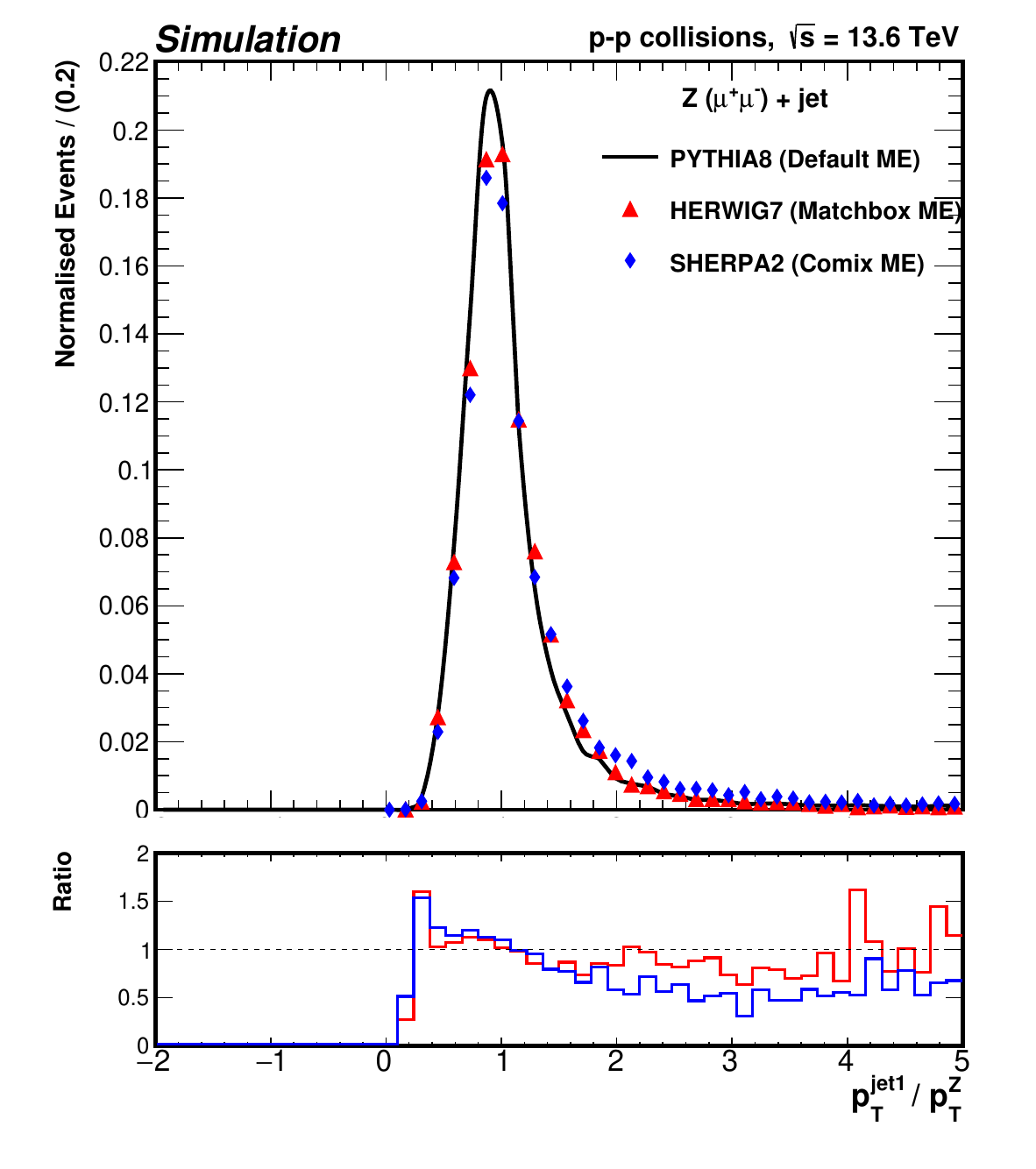}
    \caption{Normalised transverse momentum ratio, $p_{T}^{jet1}/ p_{T}^{Z}$ distributions of the leading jet and the $Z$ boson for $Z$-Jet balancing with electron channel (left) and muon channel (right) of the $Z$ boson decay with the Default ME in Pythia8, Matchbox ME in Herwig7, and Comix ME in Sherpa2. The lower panels shows the ratio of Pythia8/Herwig7 (red curve) and Pythia8/Sherpa2 (blue curve).}
    \label{ratio_pT}
\end{figure}

In order to validate the transverse momentum balance between the leading jet and $Z$ boson, we explored the back-to-back topology of the $Z$ boson and the leading jet. The angular relationship between the leading jet and $Z$ boson, as depicted by the $|\Delta\phi|$ distribution was plotted with all three event generators and is shown in Figure \ref{delta_phi}. In many standard LHC analyses, selection on $|\Delta\phi|$ variable is applied to effectively reduce backgrounds and enhance the sensitivity of the selected objects. We also present the ratio of the leading jet's transverse momentum to that of the $Z$ boson, as shown in Figure \ref{ratio_pT}.

The normalised $|\Delta\phi|$ distributions indicating the differences between the leading jet and the $Z$ boson were found to be in good agreement in all event generators with the distributions peaking $\sim$ 3.14 radian are shown in the Figure \ref{delta_phi}. Similarly, the $p_T$ ratio, $p_{T}^{jet1}/ p_{T}^{Z}$, as shown in Figure \ref{ratio_pT}, was found to be similarly modelled in all MC generators with some minor discrepancies. The $|\Delta\phi|$ distributions peaking $\sim$ 3.14 radian and the $p_T$ ratio, $p_{T}^{jet1}/ p_{T}^{Z}$ distributions peaking $\sim$ 1 indicating that the transverse momentum of the $Z$ boson and the accompanying leading jet are be roughly equal and opposite in the $Z$ + 1-jet event topology.

A cut flow table for both muon and electron decay channel analyses describing the event statistics by applying successive selections with 1 million $pp$ events is presented in Table \ref{table_cut_flow_ZJ_mu}. The estimated cross sections in millibarn ($mb$) for the $Z$+ 1-jet production are computed with the default ME in Pythia8, Matchbox ME in Herwig7 and Comix ME in Sherpa2 for both $Z\rightarrow e^+ e^-$ and $Z\rightarrow \mu^+ \mu^-$ decay channels at LO approximations and is presented in Table \ref{sigma_evt}.


\begin{table}[H]
\scriptsize
\resizebox{\textwidth}{!}{
\centering
\renewcommand{\arraystretch}{2.5}
\begin{tabular}{|c|c|c|c|c|}
\hline
 \multicolumn{2}{|c|}{\textbf{Selections}} & \textbf{Pythia8} & \textbf{Herwig7}&\textbf{Sherpa2}\\
\hline
 \multirow{3}{*}{$\mu$-channel} & Final no. of events with $p_{T}^{\mu} >$ 25 GeV/c and $|\eta_{\mu}| <$ 2.5 and $|\Delta R_{\mu, jet 1} >|$ 0.4 & 470657 & 484384 & 449524  \\
 & Final no. of events after $|\Delta \phi_{Z, jet1}| > $ 2.94 Radian cut & 459579 & 475270 & 436429 \\
& Number of events left after second jet veto & 453316 & 470903 & 426149   \\
\hline
 \multirow{3}{*}{e-channel} & Final no. of events with $p_{T}^{e} >$ 25 GeV/c and $|\eta_{e}| <$ 2.5 and $|\Delta R_{e, jet 1}| >$ 0.4 & 480264 & 493990 & 456272 \\
& Final no. of events after $|\Delta \phi_{Z, jet1}| > $ 2.94 Radian cut & 469166 & 484912 & 443496 \\
& Number of events left after second jet veto  & 463249 & 481070 & 433626  \\
\hline
\end{tabular}
}
\caption{Particle cut flow for $Z$ +1-jet events for both electron and muon decay channels of the $Z$ boson with different event generators.}
\label{table_cut_flow_ZJ_mu}
\end{table}


\begin{table}[H]
\centering
\begin{tabular}{|c|c|c|c|}
\hline
\textbf{Est. cross-section ($\sigma$) in $mb$} & \textbf{Pythia8} & \textbf{Herwig7}&\textbf{Sherpa2}\\
\hline
electron channel  & 1.952$\pm$0.0075 & 1.882$\pm$0.0037 & 1.9573$\pm$0.0032 \\
\hline
muon channel & 1.954$\pm$0.0092 & 1.884$\pm$0.0012 & 1.9572$\pm$0.0028 \\
\hline
\end{tabular}
\caption{Estimated cross-section ($\sigma$) (in mb) for $Z$+ 1-jet events with different LO event generators for 1 M events.}
\label{sigma_evt}
\end{table}

Remarkably, the cross-section ($\sigma$) values estimated through Pythia8, Herwig7, and Sherpa2 are found to be almost similar in both the electron and muon decay channels with the set of selections mentioned in Table \ref{ Cuts_table}, underlining the consistency and reliability of these event generators in simulating the hard process. Further, we notice a relatively lower values of cross sections in the Herwig7 predictions as compared to Pythia8 and Sherpa2 event generators in both the decay channels. 

\subsection{Statistical Analysis}
\label{stats}

In order to quantify the probability that the observed differences in kinematic distributions are either due to statistical fluctuations or due to the the nature of the MC simulations, we have performed Chi-square test \cite{chi_square} and Kolmogorov-Smirnov (KS) hypothesis test  \cite{kolmo} to assess the closeness or discrepancy between the results. The description of these two tests are given below.

\subsubsection{Chi - Squared Test} 

 The $\chi^2$/NDF test analysis was done on the unweighted samples of the MC using the ROOT inbuilt functions\cite{Brun:1997pa}. Different kinematics such as the $p_{T}^{Z}$, di-lepton invariant mass, $M_{\ell^{+}\ell^{-}}$, $\Delta R$ separation between the lepton candidate and the jet and the $\Delta\phi$ separation between the $Z$ boson and the leading jet are statistically analyzed. This test has been performed to quantify the degree of closeness or discrepancy in the predictions of different MC event generators. The `Pythia8' predictions were chosen as the `expected' results for Pythia8-Herwig7 and Pythia8-Sherpa2 comparison whereas `Herwig7' results were taken as the `expected' values for Herwig7-Sherpa2 comparison. The mathematical equation used for comparing the homogeneity of MC MC samples (weighted-weighted) histograms follows the below equation:
 \begin{equation}
 X^2=\sum_{i=1}^n \frac{\left(W_1 w_{2 i}-W_2 w_{1 i}\right)^2}{W_1^2 s_{2 i}^2+W_2^2 s_{1 i}^2}
 \end{equation}
 where $W_1$ and $W_2$ are the total weight of events in the first and second histogram such that $W_1$= $\sum_{i=1}^n w_{1i}$, with estimators $s_{1i}^2$ and $s_{2i}^2$ respectively. `$w_{1i}$' denotes the common weight of events of the $i^{th}$ bin in the first histogram .
A summary of $\chi^{2}$/NDF statistics for different kinematic distributions in $Z$+ 1-jet process with different event generators are reported in Table \ref{Chi-squre summary table}.

\begin{table}[H]
\centering
\tiny

\begin{tabular}{|c|c|c|c|c|c|c|c|c|}
\hline
\textbf{Kinematics} & \multicolumn{2}{|c|}{\textbf{$M_{\ell^{+} \ell^{-}}$}} & \multicolumn{2}{|c|}{\textbf{ $\Delta R_{lepton,j}$}} & \multicolumn{2}{|c|}{\textbf{$\Delta\phi_{Z, jet1}$}} & \multicolumn{2}{|c|}{\textbf{ $p_{T}^{Z}$}} \\
 \hline
 & e-channel & $\mu$-channel & e-channel & $\mu$-channel & e-channel & $\mu$-channel & e-channel & $\mu$-channel\\
\hline
Pythia8 - Herwig7 &  5.798 & 4.82 & 6.29 & 5.09 & 5.81 & 4.87 & 15.71 & 8.14\\
\hline
Pythia8 - Sherpa2 & 5.747 & 6.04 & 6.77 & 6.26 & 6.61 & 5.97 & 25.32 & 13.91\\
\hline
Herwig7 - Sherpa2 & 4.636 & 5.06 & 7.26 & 6.76 & 5.59 & 5.40 & 37.18& 36.16\\
\hline
\end{tabular}
\caption{Summary of $\chi^{2}$/NDF statistics for different kinematic distributions in $Z$ + 1-jet process with different event generators.}
 \label{Chi-squre summary table}
\end{table}


From the $\chi^{2}$/NDF test analysis, we observed that the MC-MC comparison for kinematic variables, such as $M_{\ell^{+} \ell^{-}}$, $|\Delta R_{\ell, \text{jet}1}|$, and $|\Delta \phi_{Z, \text{jet} 1}|$, yields reasonably good results of $\chi^{2}$/NDF values ranging between 4 to 6. This indicates that the predictions from the Pythia8 event generation match well with the results from Herwig7 and Sherpa2. In the case of the $p_T$ distribution of the $Z$ boson, we observed a higher statistical value (up to 37), showing the underlying differences in the event generation process, specifically in terms of the PS algorithms, parameters tuning, and assumptions in the ME calculations. The different choice of the ordering parameter for shower evolution in the Herwig7 event generator, along with variations in the UE and MPI modeling, could be the causes of the deviation in the $p_{T}^{Z}$ description from the three event generators. Furthermore, in general, all kinematic distributions seem to exhibit better agreement in the muon-decay channel of the $Z$ boson compared to the electron decay.

\subsubsection{Kolmogorov-Smirnov (KS) Test}
\begin{figure}[H]
    \centering
\includegraphics[width= .49\linewidth, height= 7cm]{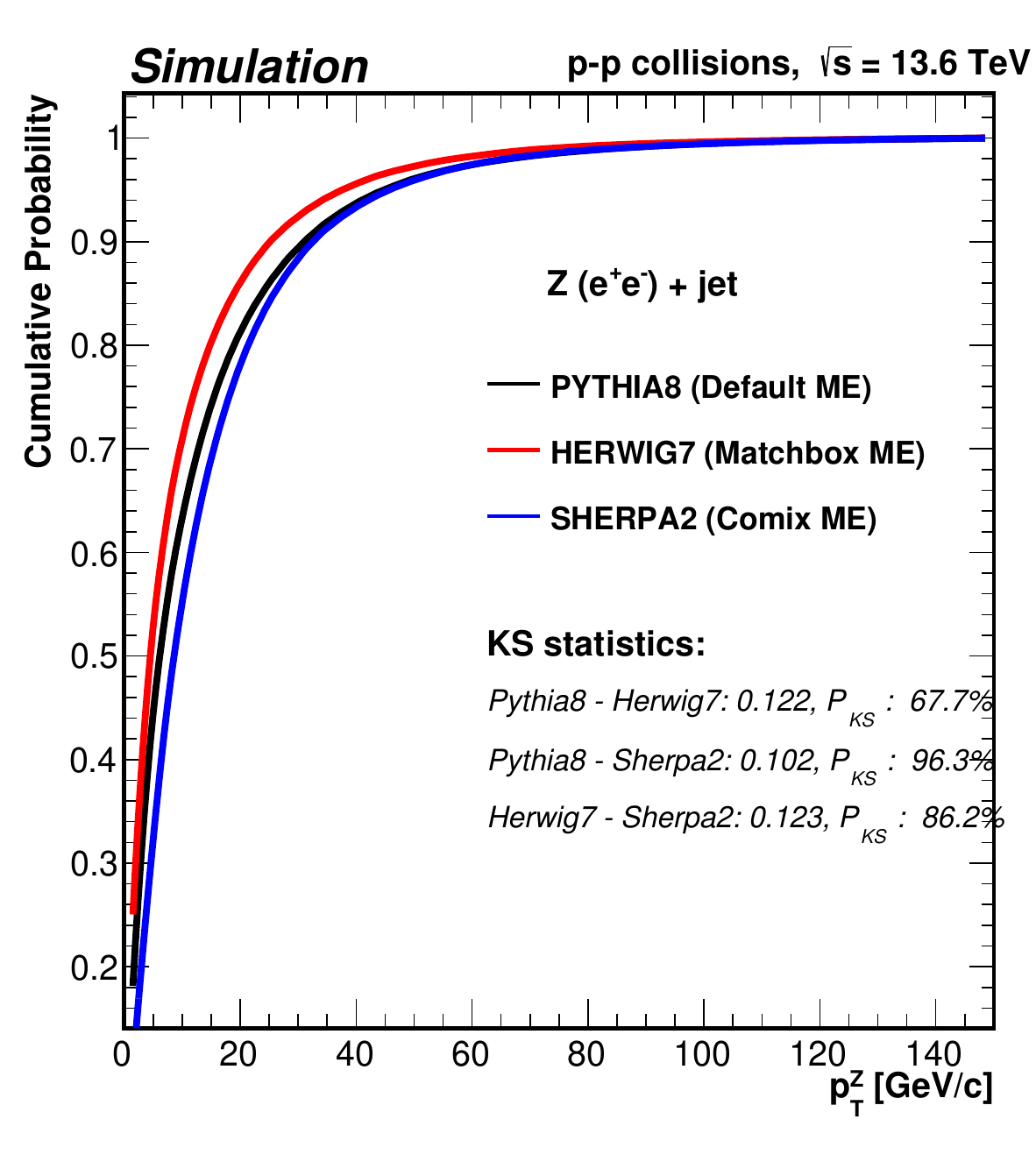}
\includegraphics[width= .49\linewidth, height= 7cm]{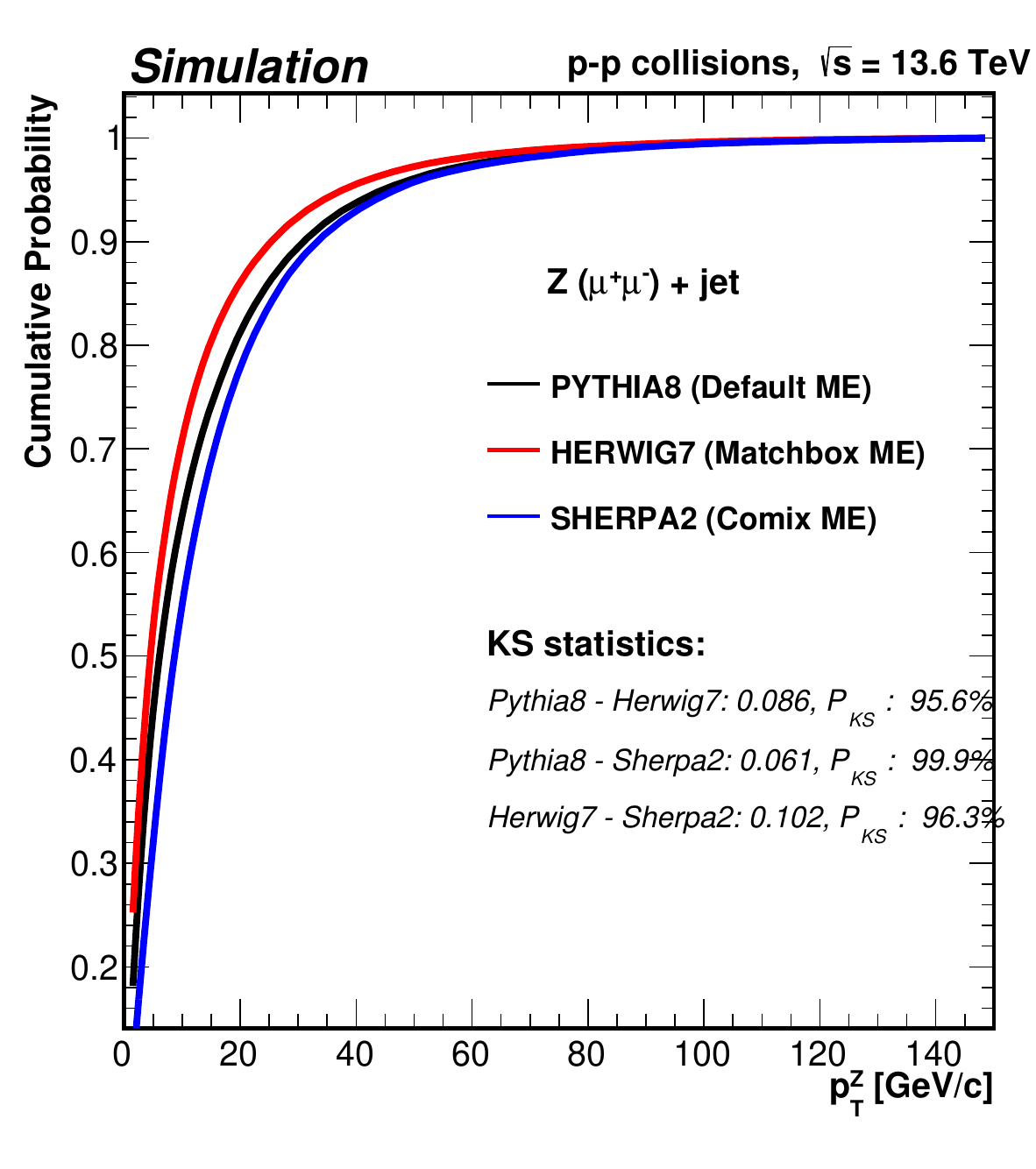}
    \caption{Cumulative Distribution Function (CDF) plots for $p_{T}^{Z}$ and Kolmogorov-Smirnov test statistics in $Z$+ 1-jet process for the electron channel (left) and muon channel (right) with different MC simulations.}
    \label{Z_pT_KS}
\end{figure}

The KS statistics were obtained by taking the maximum difference on the $y$-axis between the Cumulative Probability Function (CDF) of the $p_{T}^{Z}$ spectra. The cumulative probability of the histograms, showing the probability that a random variable takes on a value less than or equal to a specific value on y-axis, is given by $F(x) = \sum_{i=1}^{k}P_{i}$ , where $F(x)$ is the CDF at value `$x$', $P_{i}$ is the probability associated with bin `$i$' and the sum includes all bins with values less than or equal to `$x$'. Thus, the KS statistic value (D) could be obtained with $D$ = max~($F_1(x)$ - $F_2(x)$), where $F_1(x)$ and $F_2(x)$ are the cumulative probabilities at value `$x$'. The CDF plots were obtained with the help of inbuilt ROOT functions. The CDF plots and KS test statistics of CDFs for both the electron channel and muon channel are shown in Figure \ref{Z_pT_KS} and Table \ref{KS_pT} respectively. The KS Probability, $P_{KS}$ or the p-value is calculated under the null hypothesis to obtain a test statistic value as extreme as the value computed from the simulated data-sets to quantify the KS probability of agreement between the shapes of the MC distributions.

\begin{table}[H]
\centering
\begin{tabular}{|c|c|c|c|c|}
\hline
\textbf{KS statistics}& \multicolumn{2}{|c|}{\textbf{KS value}} & \multicolumn{2}{|c|}{\textbf{ $P_{KS}$}}\\
\hline
 & e-channel & $\mu$-channel & e-channel & $\mu$-channel  \\
\hline
Pythia8 - Herwig7 & 0.122 & 0.086 & 67.70\% &  95.60\%  \\
\hline
Pythia8 - Sherpa2 & 0.102 & 0.061 &  96.30\% &  99.90\%\\
\hline
Herwig7 - Sherpa2 & 0.123 & 0.102 &  86.20\% &  96.30\%\\
\hline
\end{tabular}
\caption{Summary of Kolmogorov-Smirnov test statistics for $p_{T}^{Z}$ spectra with different event generators. }
\label{KS_pT}
\end{table}


From the KS test statistics results of MC - MC summarized in Table \ref{KS_pT}, we found that the overall pattern of the $p_{T}^{Z}$ distribution showed good agreement between the results obtained from Pythia8 and Sherpa2 simulations. However, when compared with the Herwig7 results, some discrepancies were observed as evident from it's cumulative distribution shown in Figure \ref{Z_pT_KS}. This supports the result shown in Figure \ref{Z_pT}.  The best agreement was observed in the MC predictions from Pythia8 and Sherpa2 in the $Z (\mu^+ \mu^-)$ + 1-jet channel with the KS statistic value of 0.061 and KS probability 99.9\%, validating that the shapes of the two MC distributions are almost similar. The results from the KS test further supports the $\chi^{2}$ statistical values summarized in Table \ref{Chi-squre summary table}, showing a better agreement in the distributions within the muon decay channel of the $Z$ boson when compared to the electron decay channel distributions. 

\subsubsection{Percentage of Difference}
To evaluate the statistical difference between MC simulations in terms of percentage, we analyzed the di-lepton invariant mass, $M_{\ell^{+} \ell^{-}}$ and the transverse momentum, $p_{T}^{Z}$ distributions of the $Z$ boson. The percentage difference was calculated by taking the average of the absolute bin value differences such that:
\begin{equation}
      {\rm Percentage~of~difference} = \frac{1}{n} \sum_{i=1}^{n}\frac{x_{i} - y_{i}}{x_{i}} \times 100
 \end{equation}
where $n$ is the total number of bins and $x_i$ and $y_i$ are the values from the different MC datasets for the $i^{th}$ bin. The calculated values in percentage are presented in Table \ref{PD_pTZ} and Table \ref{PD_mZ} for the $p_{T}^{Z}$ and $M_{\ell^{+} \ell^{-}}$ distributions in both the electron and muon decay channels, respectively.
\begin{table}[H]
\centering

\begin{table}[H]
\centering
\begin{tabular}{|c|c|c|}
\hline
Difference  & electron channel (\%) & muon channel (\%)  \\
\hline
Pythia8 - Herwig7 & 13.70 & 13.46  \\
\hline
Pythia8 - Sherpa2 & 10.07 & 10.18 \\
\hline
Herwig7 - Sherpa2 & 14.96 & 14.86\\
\hline
\end{tabular}
\caption{Percentage of difference for $p_{T}^{Z}$ in $Z$+ 1- jet process between different MC simulations.}
\label{PD_pTZ}
\end{table}
\begin{tabular}{|c|c|c|}
\hline
 Difference & electron channel (\%) & muon channel (\%)  \\
\hline
Pythia8 - Herwig7 & 12.50 & 11.76  \\
\hline
Pythia8 - Sherpa2 & 14.72 & 14.06 \\
\hline
Herwig7 - Sherpa2 & 13.56 & 13.70\\
\hline
\end{tabular}
\caption{Percentage of difference for $M_{\ell^{+} \ell^{-}}$ in $Z$+ 1-jet process between different MC simulations.}
\label{PD_mZ}
\end{table}

We observed that the overall difference lies between 10\% and 15\% in both distributions. In the case of the $p_{T}^{Z}$ spectrum, the least difference among the three simulations is observed between Pythia8 vs Sherpa2 predictions ($\sim$10\%), followed by Pythia8 vs Herwig7 ($\sim$13\%), and Herwig7 vs Sherpa2 ($\sim$15\%). For the $M_{\ell^{+} \ell^{-}}$ spectrum, the least difference (11.7\% - 12.5\%) is observed in the case of Pythia8 vs Herwig7, followed by Herwig7 vs Sherpa2 ($\sim$13\%) and Pythia8 vs Sherpa2 ($\sim$14\%). Similar to the $\chi^2$ and KS tests, we observed that the overall closeness in the kinematic distributions were better in the muon decay channels of the $Z$ boson compared to electron channels.

\section{Other Shower Choices in Monte Carlo Event Generators}
Apart from the default showering modules  i.e. the `Simple shower'\cite{Bierlich:2022pfr} in Pythia8 , the `CS shower'\cite{Sherpa:2019gpd} in Sherpa2 and the `Default($q$-tilde) shower'\cite{Bahr:2008pv} in Herwig7, there are different alternative showering modules implemented in these event generators.
\subsection{DIRE and VINCIA showers in Pythia8}
The current version of Pythia8 comes with three different showering modules: the default `Simple shower', the `VINCIA antenna shower'\cite{Fischer:2016vfv} and the `DIRE shower'\cite{Hoche:2015sya}. Although all three shower modules are implemented by considering the dipole picture, there are different adaptations in the branching in these shower modules. In the branching of $(\Tilde{ab}) \rightarrow$($abc$), either both parents could act as the emitter and the recoiler (VINCIA) or one parent has to be the recoiler while the other emits (SIMPLE and DIRE). The alternatives VINCIA and DIRE were implemented to simulate more accurate (higher-logarithmic) and better coherent shower, whereas the simple shower operates in an improved Leading Logarithmic (LL) approximation \cite{Bierlich:2022pfr}.
 In the VINCIA shower, instead of treating the two partons differently, the whole antenna (QCD antenna represents a colour-connected parton pair which undergoes a coherent 2 $\rightarrow$ 3 branching process), treats the two pre-branching `parent' partons as a single entity. Both the participants of the dipole shares the recoil and the emission of the parton coming from the whole dipole entity. The shower evolution in the VINCIA shower is ordered in the terms of `off-shellness' which is based on a generalized version of the ARIADNE\cite{Lonnblad:1992tz} definition of transverse momentum\cite{Bierlich:2022pfr} such that for a branching $IK \rightarrow ijk$:
  \begin{equation}
 p_{\perp j}^2=\frac{\bar{q}_{i j}^2 \bar{q}_{j k}^2}{s_{\max }}
  \end{equation}
  and the off-shellness of the final state partons ($\bar{q}_{i j}$) are given by:
 \begin{equation}
     \bar{q}_{i j}^2=\left(p_i+p_j\right)^2-m_I^2=m_{i j}^2-m_I^2 
 \end{equation}
for the final-state parton $i$ and that for initial-state the partons obtained via crossing (for inital-state parton $i$) given by,
\begin{equation}
    \bar{q}_{i j}^2=-\left(p_i-p_j\right)^2+m_I^2
\end{equation}

where $p$ and $p_{\perp}$ represents the transverse momentum and its transverse component of the partons respectively, $\bar{q}$ represents the off-shellness and  $s_{\max }$ represents the maximum invariant mass squared of the system under consideration.

As of now, tree-level ME corrections are not implemented for VINCIA shower in the latest Pythia version, but one can use merging techniques using external ME providers such as Madgraph and MG5 plugins, available within the Pythia8 setup, for selecting the Born level helicities and a thorough comparison. However, to get an overview of the outcomes, we compared the Simple shower with the VINCIA shower implemented in Pythia8 at LO only.

 Another alternative for parton showering present in Pythia8 is the DIRE shower. It aims to combine aspects traditionally associated with $2\rightarrow2$ dipole (antenna) showers with features of `conventional' $1\rightarrow2$ parton showers. The goal of this hybrid is to inherit the modelling of soft-emission effects from dipole showers, while keeping an explicit association of splittings with specific collinear directions. This should allow for a comparison to the ingredients in QCD factorization theorems without any complications\cite{Bierlich:2022pfr}. In the latest version of Pythia8, the DIRE shower offers some internal functionalities for correcting the PS emissions and in our analysis, we employed those corrections for the first few splittings from the PS using ME corrections. 


\begin{figure}[H]
    \centering
\includegraphics[width= .49\linewidth, height= 7.4cm]{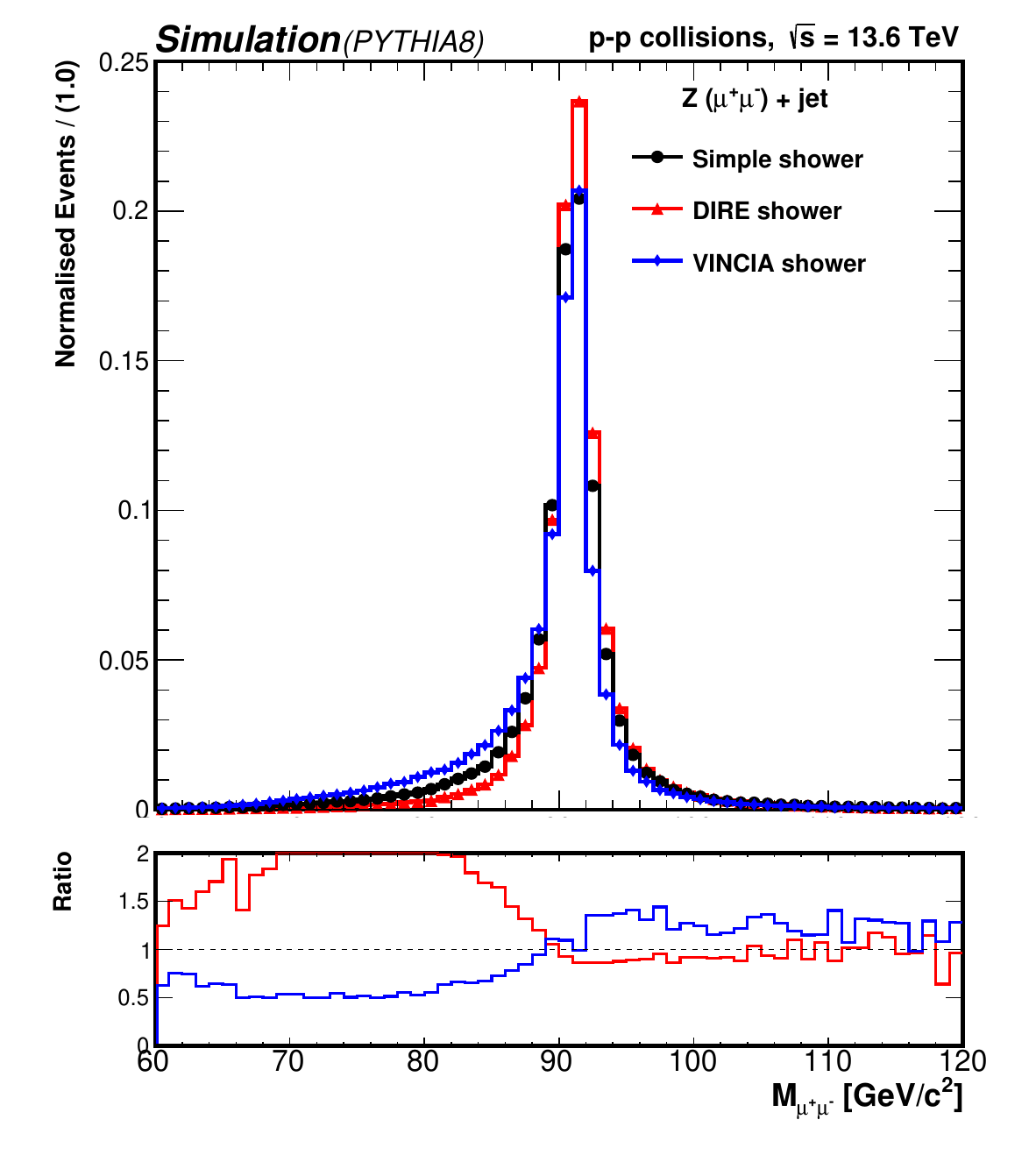}
\includegraphics[width= .49\linewidth, height= 7.4cm]{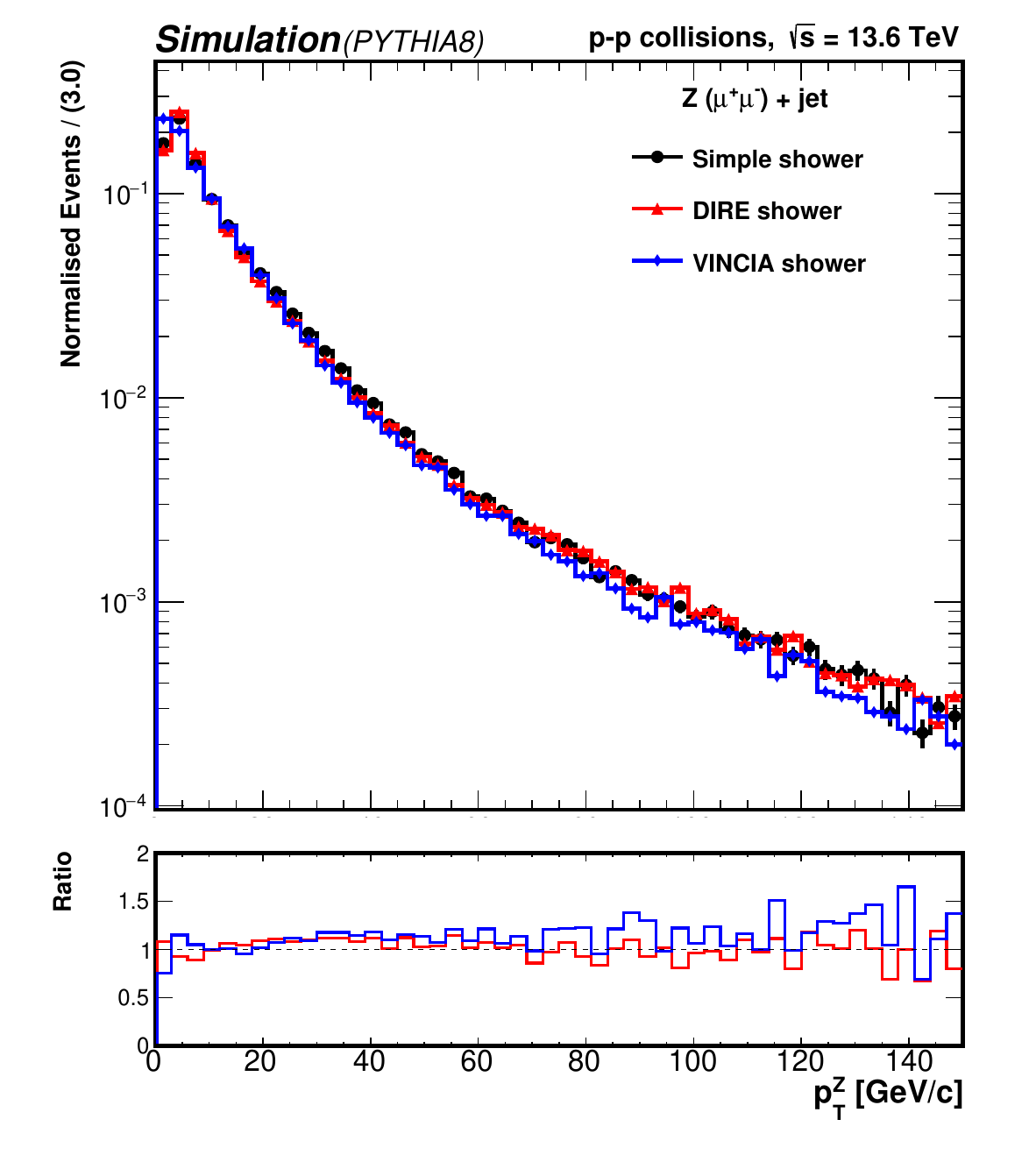} 
\caption{$Z$ boson kinematics in the $Z$+jet ($Z \rightarrow \mu^+ \mu^-$) process with different shower choices in Pythia8. The lower panel shows the ratio of Simple/DIRE (red curve) and Simple/VINCIA (blue curve).}
\label{plots_pythia_showers}    
\end{figure}

In the kinematic description of the reconstructed $Z$ boson in muon channel with different showers as shown in Figure \ref{plots_pythia_showers}, a discrepancy is observed in the distribution of the invariant mass, $M_{\mu^{+}\mu^{-}}$, while the $p_{T}^{Z}$ distribution appears to be in good agreement across all three shower choices. One possible reason for this discrepancy in the invariant mass distribution could be the treatment of finite-width effects, called interleaved resonance decays\cite{Bewick:2023tfi}, which are implemented differently in these shower algorithms. VINCIA’s shower modules are fully interleaved with Pythia’s treatment of MPI, in the same manner as for the simple-shower model\cite{Bierlich:2022pfr}, which can show some subtle changes in the invariant mass distribution when compared with the results from the DIRE shower.

\subsection{Dipole Shower in Herwig7}
Herwig7 comes with two parton showering modules: the default angular-ordered ($q$-tilde) shower and the dipole shower (Catani–Seymour) \cite{Platzer:2009jq,CS_1,Catani:2002hc}. We used the default $q$-tilde shower (also called the angular-ordered shower) for the comparison of results reported in Section \ref{results}, in which the shower is simulated as a series of individual emissions, and only the shower variables ($\Tilde{q},z, \phi$) are calculated for each emission. Once there is no phase space available for further branching, the external particles are taken to be on-shell and the physical momenta are reconstructed.

In the dipole shower algorithm, the shower evolves from a hard sub-process by sorting the colored partons attached to the hard sub-process through the color flow information in the large-$N_c$ limit\cite{Platzer:2009jq, Platzer:2011bc}. The partons in each singlet are sorted such that colour connected partons are located at neighbouring positions, when representing the singlet group of partons as a sequence. These singlet sequences are referred as the dipole chains which forms a `dipole' from each pair of the subsequent partons in a singlet sequence and the hard scale is determined for each parton in each dipole. The dipole chains are removed if they cannot further branch. For the dipoles, which can emit partons, one candidate has to face the `recoil' of the `emitter' candidate and the showering looks like as $(i,j)\rightarrow (i',k,j)$ for a $(i,j)$ pair. The full description of the shower implementation in Herwig7 and its specifications are given in Ref. \cite{Platzer:2011bc}.
\begin{figure}[H]
    \centering
\includegraphics[width= .49\linewidth, height= 7.4cm]{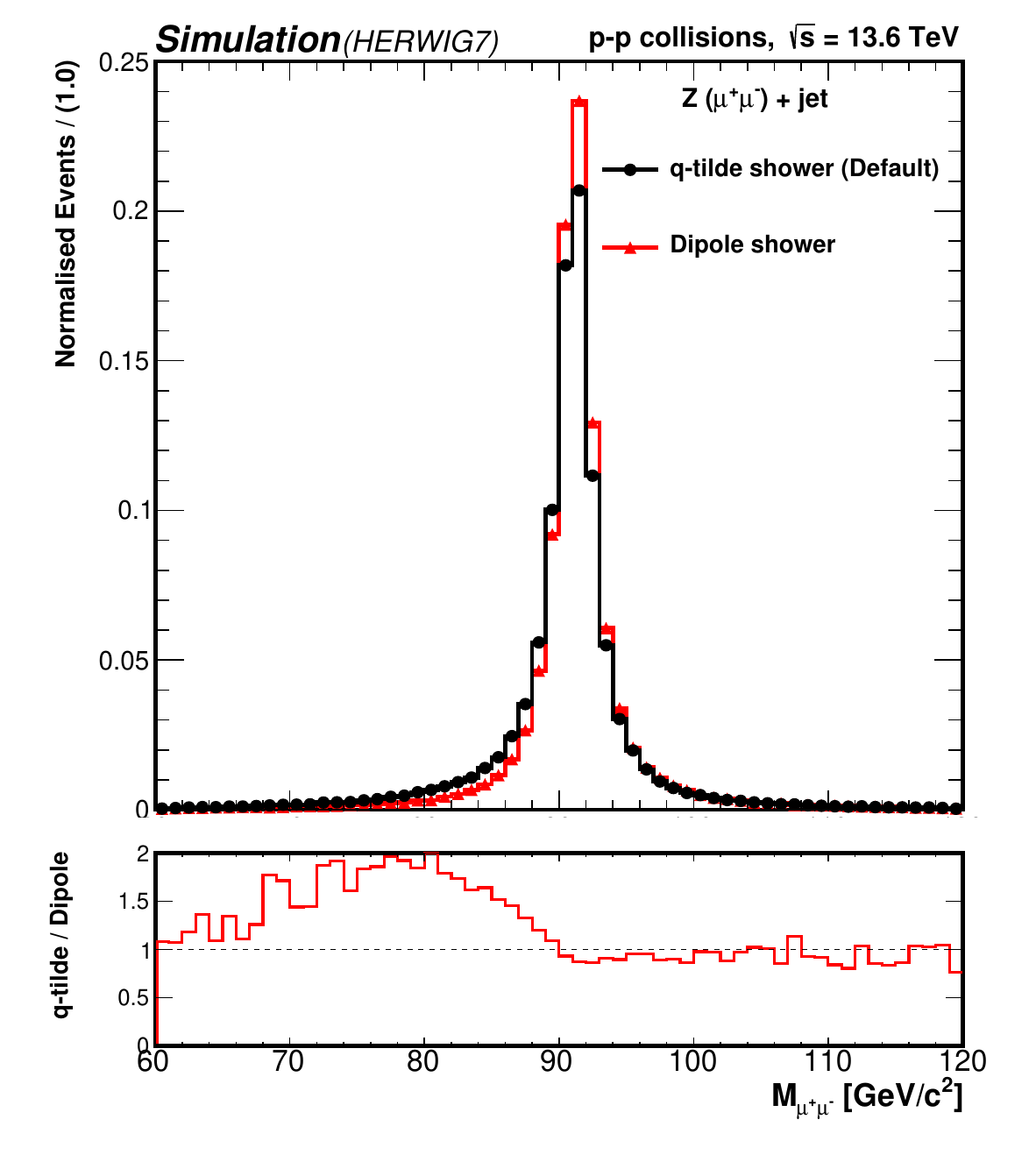}
\includegraphics[width= .49\linewidth, height= 7.4cm]{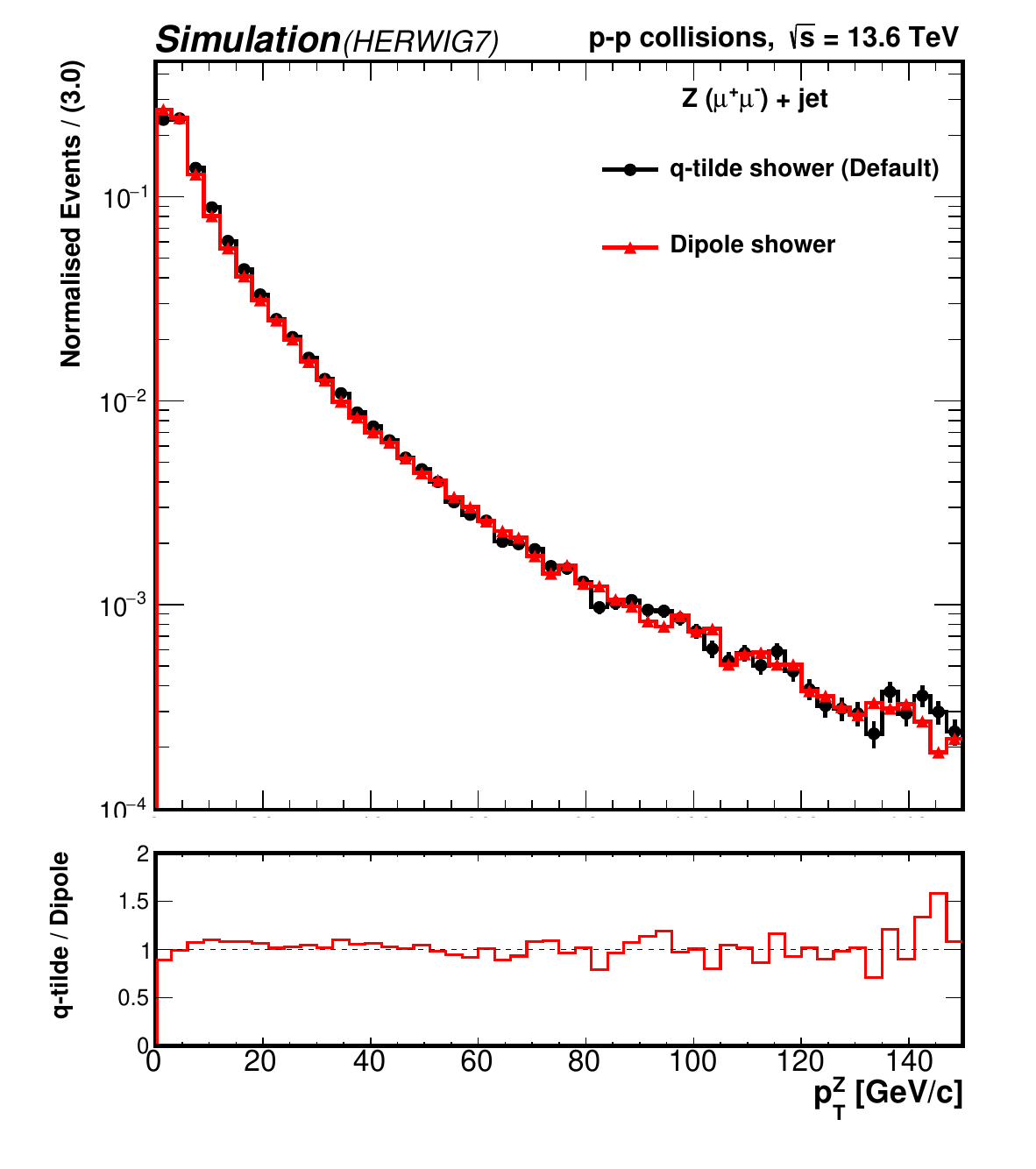}
\caption{ $Z$ boson kinematics in the $Z$+ 1-jet ($Z \rightarrow \mu^+ \mu^-$) process with different shower choices in Herwig7. The lower panel shows the ratio of $q$-tilde/Dipole showers.}    
\label{plots_shower_herwig}
\end{figure}



From the kinematic descriptions of the reconstructed $Z$ boson in muon channel shown in Figure \ref{plots_shower_herwig}, we did not notice any significant difference in $p_{T}^{Z}$ spectrum predicting the output at LO accuracy. However, some level of discrepancy was observed in modeling the $M_{\mu^{+}\mu^{-}}$ spectrum especially in the lower mass region of $Z$ boson.








\subsection{DIRE shower in Sherpa2}
In the latest version of Sherpa, there are two showering modules available for the parton showering: the default CS shower and the DIRE shower. The CS shower relies on the factorization of real-emission MEs in the CS subtraction framework \cite{Sherpa:2019gpd} and there are four general types of CS dipole terms that capture the complete infrared singularity structure of NLO QCD amplitudes\cite{Gleisberg:2008ta}. In the DIRE model, the dipole-like picture of the evolution (emitter and recoiler) can be re-interpreted as a dipole picture in the soft limit and the splitting functions are then modified to satisfy the sum rules in the collinear limit\cite{Gleisberg:2008ta}.

\begin{figure}[H]
\centering
\includegraphics[width= .49\linewidth, height= 7.4cm]{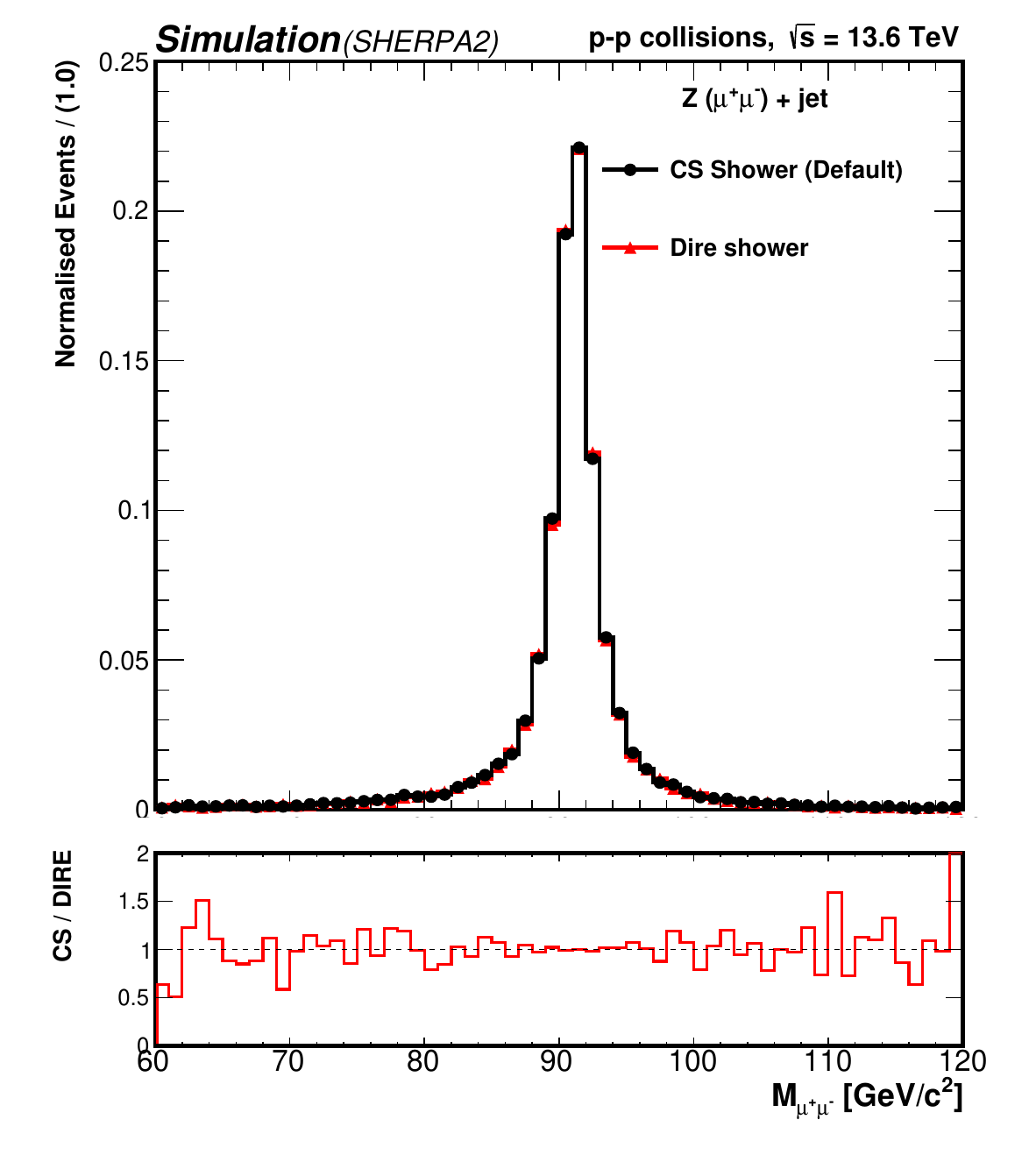}
\includegraphics[width= .49\linewidth, height= 7.4cm]{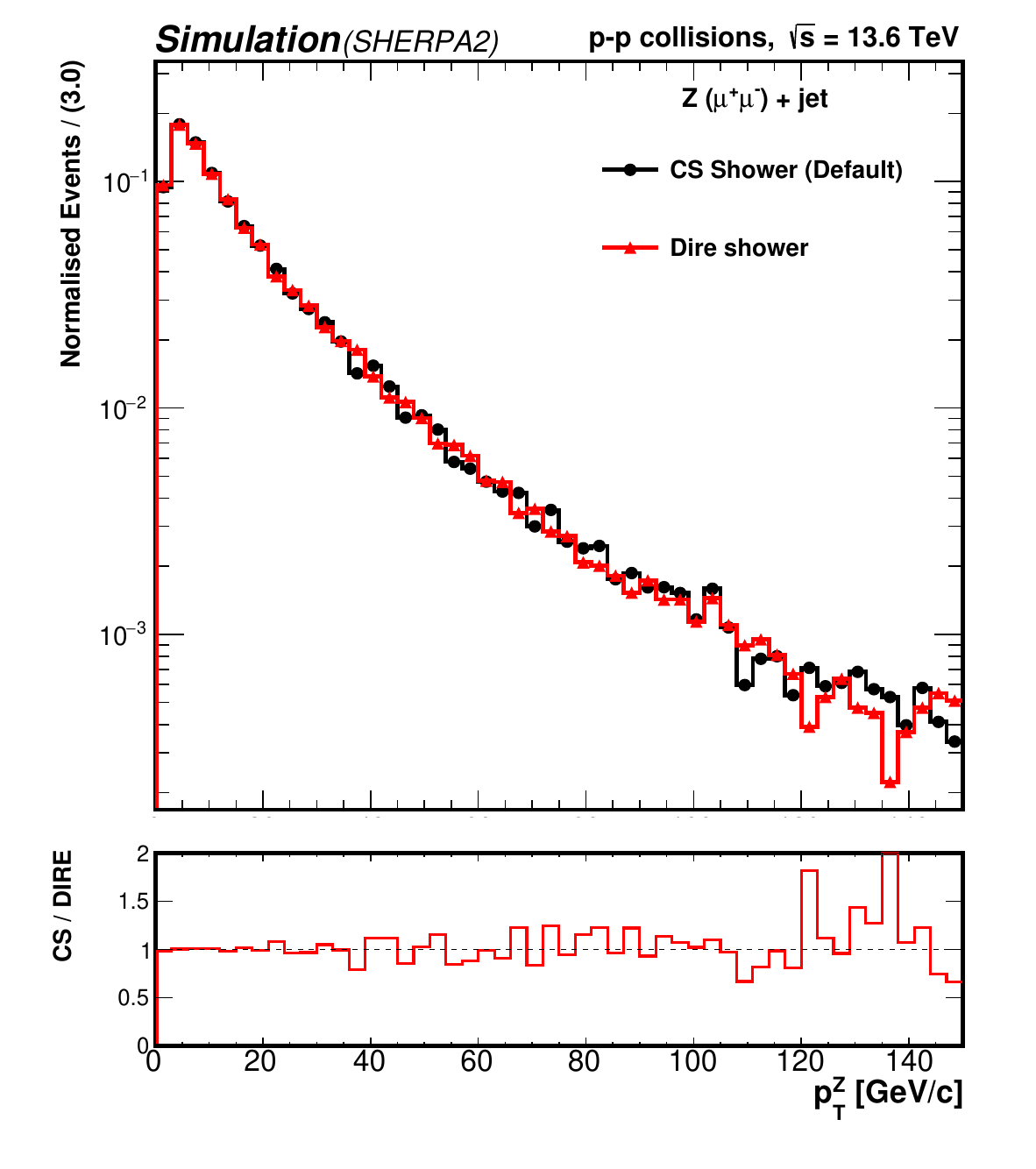}

\caption{$Z$ boson kinematics in the $Z$+ 1-jet ($Z \rightarrow \mu^+ \mu^-$) process with different shower choices in Sherpa2. The lower panel shows the ratio CS / DIRE showers.}
\label{plots_shower_sherpa}
\end{figure}

    
A comparison of dimuon invariant mass, $M_{\mu^{+}\mu^{-}}$, and transverse momentum of $Z$ boson, $p_{T}^{Z}$ distributions with CS and DIRE showers of Sherpa2 is presented in Figure \ref{plots_shower_sherpa}. The results demonstrate very good agreement in kinematics of $Z$ boson between the two types of showering modules available in Sherpa2.



\section{Summary}

In this paper we report a comprehensive study of the production of the $Z$ boson in association with a jet using LO event generators such as Pythia8, Herwig7, and Sherpa2. Simulated proton-proton events at a center-of-mass energy of $\sqrt{s} = 13.6$ TeV at the LHC are investigated. The study focuses on two decay channels of the $Z$ boson, namely $e^{+}e^{-}$ and $\mu^{+}\mu^{-}$, each with a specific set of kinematic selections. The analysis of $Z$+jet events not only plays a crucial role in SM measurements but also contributes significantly to the search for new physics at the LHC. Being a prevalent background process, the study of the kinematics and event topology of the $Z$+ 1-jet process in this paper provides valuable insights for testing predictions from perturbative QCD and understanding the underlying physics events. This study on simulations is particularly useful for theoretical predictions, Monte Carlo tuning, and consideration in experimental analyses at LHC.

The internal ME corrections are employed in three of the event generators, and their predictions for the kinematic distributions in $Z$+ 1-jet production are investigated. Several kinematic variables of the $Z$ + 1-jet process, such as lepton transverse momentum $p_T^{\ell}$, dilepton invariant mass, $M_{\ell^{+}\ell^{-}}$, transverse momentum of the $Z$ boson, $p_T^{Z}$, jet multiplicity, $N_{\text{Jet}}$, leading jet transverse momentum, $p_{T}^{\text{jet1}}$, the angular separation between the leading jet and leptons, $\Delta R_{\text{jet1,}\ell}$, the azimuthal angle between the leading jet and $Z$ boson, $\Delta \Phi_{\text{jet1, Z}}$, and the $p_T$ ratio of the leading jet and $Z$ boson, $p_T^{\text{jet1}}/p_T^{Z}$, are investigated. 

From this study we observed that the overall shapes of the kinematic distributions such as $p_T^{\ell}$, $M_{\ell^{+}\ell^{-}}$ and $p_T^{Z}$ are similarly modeled by Pythia8 (with default ME) and Sherpa2 (with Comix ME) predictions. However, Herwig7 (with Default ME) exhibits a larger discrepancy with respect to both Pythia8 and Sherpa2, prompting an exploration of other ME choices such as Matchbox ME within Herwig7. The incorporation of Matchbox ME within Herwig7 significantly improved the performance of the kinematic variables with Herwig7 simulations, aligning the $p_T^{\ell}$, $M_{\ell^{+}\ell^{-}}$ and $p_T^{Z}$ distributions more closely with those predicted by Pythia8 (with default ME) and Sherpa2 (with Comix ME). Similarly, the matching of the distributions of the $p_{T}^{\text{Jet1}}$, $\Delta R_{\text{Jet1,} \ell}$, $\Delta \Phi_{\text{Jet1, Z}}$, and $p_T^{\text{Jet1}}/p_T^{Z}$ with the application of Matchbox ME in Herwig7 improved and becomes more close with Pythia8 and Sherpa2.  

To compute the degree of closeness or discrepancy in modeling in some of the important kinematic variables of the $Z$ boson and jets among the three event generators, $\chi^2$/NDF and KS tests are conducted through a comparison between two distinct MC event sets such as Pythia8 vs Herwig7, Pythia8 vs Sherpa2 and Herwig7 vs Sherpa2. The $\chi^2/NDF$ test is conducted on variables such as $p_T^{Z}$, $M_{\ell^{+}\ell^{-}}$, $\Delta R_{\text{Jet1,} \ell}$, and $\Delta \Phi_{\text{Jet1, Z}}$, whereas the KS test is specifically performed on the $p_T^{Z}$ variable. The KS test is done to calculate the KS statistics and probability (p-value). Based on the $\chi^2$/NDF analysis, we observed that all kinematic distributions except the $p_{T}^{Z}$ spectrum were nearly similarly modeled across all three event generators, with $\chi^2$/NDF values ranging between 4 and 6 in both electron and muon decay channel analyses. The most substantial disagreement in the case of the $p_T^{Z}$ spectrum in the electron channel analysis was observed between Herwig7 and Sherpa2, with a higher $\chi^2/NDF$ value of approximately 37. 

In the KS test, a close trend of the $p_{T}^{Z}$ spectrum was observed in the Pythia8 and Sherpa2 predictions, where the KS test statistic value was found to be around 0.061 corresponding to the KS probability (p-value) of 99.9\%. However, larger KS test statistics values of around 0.12 were observed when we compare Herwig7 vs Sherpa2 and Pythia8 vs Herwig7 indicating a relatively larger discrepancy between their predicted distributions, possibly because of the different choice of evolution parameter in their PS algorithm ($p_T$ based variable in Pythia8 and Sherpa2 and an angular based parameter in Herwig7). Overall, it was noticed that all the kinematic distributions in the muon channel analysis better match in different event generators compared to the electron channel distributions.

 Furthermore, the possible alternative showering modules provided within these MC event generator frameworks are explored and compared their predictions with the default settings. It was found out that the three showering modules in Pythia8 such as Simple shower, VINCIA shower and DIRE shower give similar description of the $p_{T}^{Z}$ spectrum with a slight discrepancy in the $M_{\ell^{+}\ell^{-}}$ distribution of the resonance particle ($Z$ boson in our case) when the DIRE shower results are compared with the results from the default Simple shower. A similar deviation was observed in the comparison plots of the $M_{\ell^{+}\ell^{-}}$ distribution when the two shower modules present in Herwig7 i.e, the dipole shower and the default $q$-tilde shower were compared. The alternative shower module present in Sherpa2 (DIRE shower) however reproduces the $p_{T}^{Z}$ and $M_{\ell^{+}\ell^{-}}$ descriptions almost similar as with it's corresponding default parton shower (CS shower).

\acknowledgments 
Bibhuti Parida acknowledges the Science and Engineering Research Board (SERB), Government of India, for the grant received under the Accelerate Vigyan scheme (File No: AV/VRI/2022/0228). This grant supported the organization of student internships during $\rm 6^{th}$ February to $\rm 13^{th}$ March 2023 at Amity University Uttar Pradesh, Noida, where part of this work has been done. Dharmender acknowledges the internship opportunity and for the numerous discussions throughout the aforementioned event.






\bibliographystyle{JHEP}
\bibliography{biblio.bib}




\end{document}